\documentclass[trackchanges,twocolumn]{aastex7}

\usepackage{amsmath}
\usepackage{multirow}

\makeatletter
\newcounter{subsubsubsection}[subsubsection]

\newcommand\l@subsubsubsection{\@dottedtocline{4}{7em}{4em}}
\makeatother

\begin{document}

\title{PANORAMIC: The Dawn of Massive Quiescent Galaxies I. \\
Number Density and Cosmic Variance from 1000 arcmin$^2$ NIRCam Imaging}

\correspondingauthor{Zhiyuan Ji}
\email{zhiyuanji@arizona.edu}

\author[0000-0001-7673-2257]{Zhiyuan Ji}
\affiliation{Steward Observatory, University of Arizona, 933 N. Cherry Avenue, Tucson, AZ 85721, USA}
\email{zhiyuanji@arizona.edu}

\author[0000-0003-2919-7495]{Christina C. Williams}
\affiliation{NSF–DOE Vera C. Rubin Observatory/NSF NOIRLab, 950 N. Cherry Ave., Tucson, AZ 85719, USA}
\affiliation{Steward Observatory, University of Arizona, 933 N. Cherry Avenue, Tucson, AZ 85721, USA}
\email{christina.williams@noirlab.edu}

\author[0000-0002-2517-6446]{Peter Behroozi}
\affiliation{Steward Observatory, University of Arizona, 933 N. Cherry Avenue, Tucson, AZ 85721, USA}
\email{behroozi@arizona.edu}

\author[0000-0001-8928-4465]{Andrea Weibel}
\affiliation{Department of Astronomy, University of Geneva, Chemin Pegasi 51, 1290 Versoix, Switzerland}
\email{Andrea.Weibel@unige.ch}

\author[0000-0002-8896-6496, gname=Christian Kragh, sname=Jespersen]{Christian Kragh Jespersen}
\affiliation{Department of Astrophysical Sciences, Princeton University, Princeton, NJ 08544, USA}
\email{cj1223@princeton.edu}

\author[0000-0001-5851-6649]{Pascal A. Oesch}
\affiliation{Department of Astronomy, University of Geneva, Chemin Pegasi 51, 1290 Versoix, Switzerland}
\affiliation{Cosmic DAWN Center, Niels Bohr Institute, University of Copenhagen, Jagtvej 128, K\o benhavn N, DK-2200, Denmark}
\email{Pascal.Oesch@unige.ch}

\author[0000-0001-5063-8254]{Rachel Bezanson}
\affiliation{Department of Physics and Astronomy and PITT PACC, University of Pittsburgh, Pittsburgh, PA 15260, USA}
\email{rachel.bezanson@pitt.edu}

\author[0000-0001-7160-3632]{Katherine E. Whitaker}
\affiliation{Department of Astronomy, University of Massachusetts, Amherst, MA 01003, USA}
\affiliation{Cosmic DAWN Center, Niels Bohr Institute, University of Copenhagen, Jagtvej 128, K\o benhavn N, DK-2200, Denmark}
\email{kwhitaker@astro.umass.edu}

\author[0000-0002-5612-3427]{Jenny E. Greene}
\affil{Department of Astrophysical Sciences, 4 Ivy Lane, Princeton University, Princeton, NJ 08540}
\email{jgreene@astro.princeton.edu}

\author[0000-0003-2680-005X]{Gabriel Brammer}
\affiliation{Cosmic Dawn Center (DAWN), Denmark}
\affiliation{Niels Bohr Institute, University of Copenhagen, Jagtvej 128, K{\o}benhavn N, DK-2200, Denmark}
\email{gabriel.brammer@nbi.ku.dk}

\author[0000-0001-8460-1564]{Pratika Dayal}
\affiliation{Canadian Institute for Theoretical Astrophysics, 60 St George St, University of Toronto, Toronto, ON M5S 3H8, Canada}
\affiliation{David A. Dunlap Department of Astronomy and Astrophysics, University of Toronto, 50 St. George Street, Toronto, Ontario, M5S 3H4, Canada}
\affiliation{Department of Physics, 60 St George St, University of Toronto, Toronto, ON M5S 3H8, Canada}
\email{pdayal@cita.utoronto.ca}

\author[0000-0002-2057-5376]{Ivo Labb\'e}
\affiliation{Centre for Astrophysics and Supercomputing, Swinburne University of Technology, Melbourne, VIC 3122, Australia}
\email{ilabbe@swin.edu.au}

\author[0000-0003-0415-0121]{Sinclaire M. Manning}
\affiliation{Department of Astronomy, University of Massachusetts, Amherst, MA 01003, USA}
\email{smanning.astro@gmail.com}

\author[0000-0002-5104-8245]{Pierluigi Rinaldi}
\affiliation{Space Telescope Science Institute, 3700 San Martin Drive, Baltimore, Maryland 21218, USA}
\email{prinaldi@stsci.edu}

\author[0000-0003-1207-5344]{Mengyuan Xiao}
\affiliation{Department of Astronomy, University of Geneva, Chemin Pegasi 51, 1290 Versoix, Switzerland}
\email{Mengyuan.Xiao@unige.ch}

\author[0000-0001-6454-1699]{Yunchong Zhang} 
\affiliation{Department of Physics and Astronomy and PITT PACC, University of Pittsburgh, Pittsburgh, PA 15260, USA}
\email{yunchongzhang@pitt.edu}

\begin{abstract}

We measure the number density and field-to-field variance of massive quiescent galaxies at $z\sim3$--8 using the JWST/NIRCam pure-parallel imaging survey PANORAMIC together with archival observations, covering an area of 0.28 deg$^2$ ($\sim1000$ arcmin$^2$) in at least six filters. We identify quiescent galaxy candidates at $z\gtrsim3$ with $M_\ast \gtrsim 10^{10}\,M_\odot$, comprising 101 galaxies in a gold sample of high-confidence candidates and 137 in a more inclusive silver sample. We measure their evolving comoving number density, finding $(1.5$ vs. $3.1)\times10^{-5}\,\mathrm{Mpc}^{-3}$ at $z=3$--4 for the gold and silver samples, respectively, and a decline by more than a factor of 20 by $z\sim6$. Comparisons with empirical models and cosmological simulations show that widely used frameworks underpredict the abundance of massive quiescent galaxies at $z\gtrsim4$ by $\gtrsim1$ dex, indicating that current implementations of early star formation, feedback, and quenching do not produce enough early quenched systems. With 34 independent sightlines, we present the first direct empirical measurement of field-to-field variance for quiescent galaxies at $z>3$, finding a high cosmic variance of $\sigma_{\rm CV}\approx0.7\pm0.3$. This exceeds predictions from abundance-matched mock catalogs, suggesting that early quiescent galaxies are more strongly clustered, and more likely to be found near one another or in more biased regions, than expected in current galaxy-formation models. Any successful model for the emergence of early massive quiescent galaxies must reproduce both their abundance evolution and their imprint on the large-scale distribution.

\end{abstract}

\section{Introduction} 
\label{sec:intro}

The formation of massive quiescent galaxies remains a major theoretical challenge in astrophysics. Stellar archaeology indicates that their stars formed in an extreme and rapid burst, followed by rapid truncation of star formation before $z>3$ (\citealt{Thomas2005,Thomas2010}). Such early assembly implies star-formation efficiencies in massive galaxies that were more than an order of magnitude higher in the first Gyr \citep{Xiao2024, Xiao2025, deGraaff2025Nat} than those typical today \citep{SaintongeCatinella2022}. Meanwhile, their lower-redshift descendants remain physically distinct from the bulk of the star-forming population, typically exhibiting compact, centrally concentrated morphologies and passive evolution over billions of years \citep{vandokkum2008, Hopkins2010,Grudic2019}. 

Decades of work at $z\lesssim3$ have sought to understand how mass, morphology, and local environment shape the quenching of these systems \citep[e.g.][]{Peng2010, Peng2015, Ji2018, Jespersen2025b}, yet no complete theoretical picture has emerged. Massive quiescent galaxies exhibit a diversity of quenching timescales \citep{Carnall2020, Belli2019}, likely linked to the disappearance of cold molecular gas and dust \citep{Williams2021,Whitaker2021, Bezanson2022}. But the empirical signatures of the relevant quenching mechanisms remain elusive, perhaps because they are short-lived and most visible only during the quenching phase itself \citep{Park2023, Belli2024, Bugiani2025}. As time passes, other unrelated processes such as merging and rejuvenation \citep[e.g.,][]{Woodrum2022,Ji2022, Ji2023, Suess2023, Suess2025} complicate the forensic reconstruction  of quenching mechanisms. These difficulties motivate direct searches for massive quiescent galaxies during the epoch when they first formed and quenched, i.e. $z\gtrsim3$. Further, studies of quiescent galaxies at $z>3$ present a complementary test of the early formation histories inferred from lower-redshift quenched systems.

Pre-JWST, candidate galaxies were confirmed with ground-based facilities out to $z_{\rm spec}=3.7$, revealing old stellar populations with extremely early formation times \citep{Glazebrook2017,Kriek2016,Schreiber2018,AntwiDanso2025}. However, the limited wavelength coverage from the ground ($\lambda<2.5\,\mu$m) prevents accurate stellar population modeling beyond $z>4$, where longer-wavelength spectroscopy is needed to measure the Balmer break and age-sensitive absorption features. The launch of JWST \citep{Gardner2023} transformed this field by enabling efficient spectroscopic confirmation of quiescent galaxies at $z>3$. JWST has now confirmed quiescent galaxies at $z\sim4$--5 \citep{Nanayakkara2025,Carnall2024,Carnall2024excel,deGraaff2025Nat,Ito2026,Kakimoto2024,Zhang2025,Baker2026,Hamadouche2026}, with one known example at $z=7.3$ \citep{Weibel2025}. The improvement in measured formation and quenching timescales enabled by JWST is critical to the development of theories about their formation pathways.

A key missing piece of information is how the abundance of massive quiescent galaxies evolves with redshift, since this provides a direct statistical test of cosmological models for rapid growth and quenching. 
Unfortunately, massive ($M_\ast \gtrsim 10^{10}\,M_\odot$) quiescent galaxies at $z>3$ are extremely rare: current estimates predict only 1–2 detections at $z>4$ in a typical $\sim200$ arcmin$^2$ pencil-beam survey \citep[e.g.,][]{Valentino2023}.
Multiple JWST studies have reported abundances that exceed simulation predictions, but these results are based on only 2--6 independent sightlines \citep{Valentino2023, Carnall2022, Long2024, Zhang2025, Alberts2024, Merlin2025, Stevenson2026, Russell2025, Baker2025b}.
Taken together, these studies suggest that current measurements at $z>3$ are strongly affected by field-to-field variance, likely reflecting strong clustering of this highly biased population \citep[e.g.][]{Valentino2023}. Robust constraints on their abundance therefore require surveys that probe both substantially larger area and many independent sightlines. Yet, JWST is inefficient for contiguous wide-field imaging because of large slews, overheads, and the small field of view of the near-infrared camera \citep[NIRCam;][]{Rieke2023a}.

Fortunately, JWST can obtain NIRCam imaging in pure parallel mode, enabling extremely efficient wide-area imaging while naturally sampling independent sightlines, the most effective strategy for characterizing the abundance and spatial distribution of rare, highly biased populations \citep[e.g.,][]{Steinhardt2021, Trapp2022}. This strategy has strong precedent from HST parallel programs such as BoRG and HIPPIES \citep[][]{Trenti2011, Yan2011,Bradley2012,Calvi2016}, which surveyed $\sim0.4$ deg$^2$ in 3--4 filters, discovered bright $z\sim6$--10 galaxies too rare for pencil-beam surveys, and reduced cosmic variance by observing more than 300 independent sightlines \citep[e.g.,][]{Morishita2018,Bridge2019,RojasRuiz2020,Morishita2021,RobertsBorsani2022,Bagley2024}. Although limited in ancillary data, these HST parallel surveys remain an important legacy dataset: some of their brightest $z\sim6$--9 Lyman-break galaxies have now been spectroscopically confirmed with JWST \citep{RobertsBorsani2025, RojasRuiz2025a}, and the parallels provided essential context for interpreting the unexpectedly high abundance of bright galaxies at $z>10$ \citep{RojasRuiz2025b}.

NIRCam pure parallels are significantly more powerful. Owing to the dichroic design, they provide 2--3$\times$ the filter coverage of HST parallels \citep{Williams2025, Morishita2025}. To date they have nearly doubled the area with $\geq6$ NIRCam filters (required for robust $z>4$ selection) and increased the number of blank-field independent sightlines by $\gtrsim6\times$. These capabilities enable new constraints on the abundance and clustering of quiescent galaxies at $z>4$, previously limited by small areas and cosmic variance. These data provide the first characterization of the clustering of this highly biased population, informing the galaxy–halo connection \citep[e.g.,][]{Robertson2010, Weibel2025b} and the role of large-scale environment in early quenching. Such constraints are essential for interpreting high abundances in rare, small-area samples \citep{Jespersen2025}, and clustering measurements are key to understanding quiescent galaxy evolution in a cosmological context \citep{Jespersen2025b}.

In this paper, we present results using pure parallel imaging from the PANORAMIC Survey \citep{Williams2025} combined with archival imaging from JWST Cycles 1-3 programs. The data comprises a total of 1000 arcmin$^2$ of NIRCam imaging across an unprecedented 34 independent sightlines.  The paper is organized as follows.  In Section \ref{sec:data} we describe the pure parallel and archival imaging data. In Section \ref{sec:selection} we describe the selection of our $z>3$ quiescent galaxy sample. In Section \ref{sec:n}  we provide our measurement of the abundance evolution of quiescent galaxies at $z>3$, utilizing 34 independent sightlines (a sixfold increase over previous studies, enabling the most robust mitigation of cosmic variance to date). Finally, in Section \ref{sec:cv}, we provide a first characterization of the clustering properties of $z>3$ quiescent galaxies by measuring the empirical field-to-field variance, which can be compared to simulations to gain insight into the underlying dark matter halo population that quiescent galaxies trace. We jointly interpret our abundance and clustering constraints, and their implications for the early quenching of massive galaxies, in Section \ref{sec:disc}.

Throughout this paper, we use “environment” in a purely descriptive sense, to denote the large-scale spatial context traced by cosmic variance, rather than any specific quenching mechanism. We adopt the AB magnitude system and the $\Lambda$CDM cosmology with \citealt{Planck2020} parameters, i.e., $\Omega_m = 0.315$ and $\rm{h = H_0/(100 km\,s^{-1}\,Mpc^{-1}) = 0.673}$.

\begin{figure*}
    \includegraphics[width=1\textwidth]{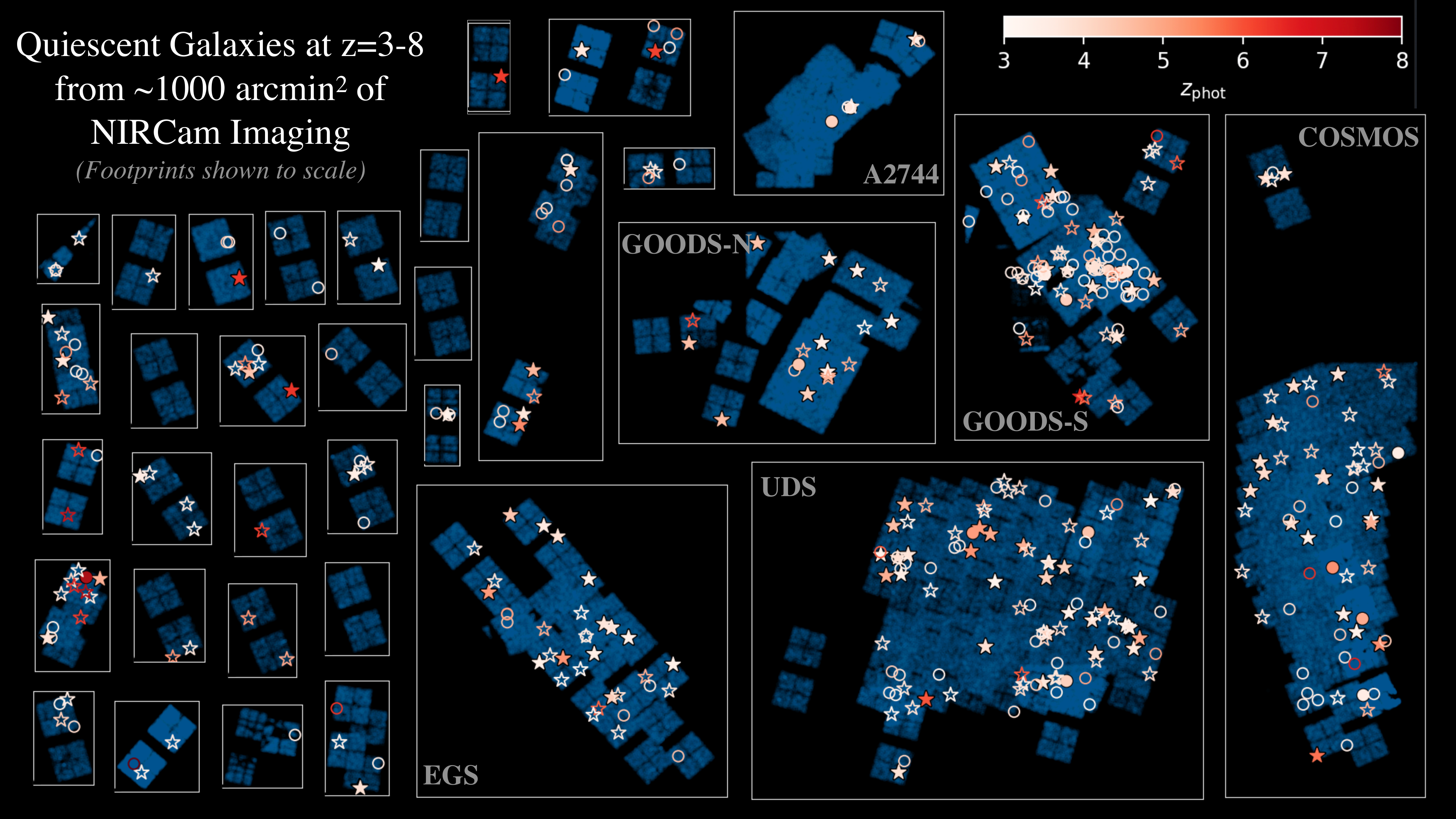}
    \caption{Quiescent galaxies at $z\sim3$–8 (a total of 406) identified in this work from $\sim$1000 arcmin$^2$ of JWST/NIRCam imaging. Galaxies in the gold and silver samples are shown as filled and open symbols, respectively (see Section \ref{sec:selection}). Massive systems with $M_*\geq10^{10}M_\odot$ -- the main focus of this study (101 in the gold sample and 137 in the silver sample) -- are shown as stars, while less massive ones are shown as circles. Footprints separated by $<1^\circ$ are combined. The single NIRCam pointings come from the pure-parallel program PANORAMIC \citep{Williams2025}. Because our  selection requires at least 6-filter NIRCam coverage, in the COSMOS field only the regions around the PRIMER survey are included. The strong impact of cosmic variance (i.e., spatial clustering) on the distribution of $z\geq3$ quiescent galaxies is visually evident in this plot.}
    \label{fig:footprints}
\end{figure*}

\section{Data}\label{sec:data}

This work is based on publicly available imaging from JWST/NIRCam and HST. To ensure robust photometric redshift constraints, we restrict our analysis to regions with coverage in at least 6 NIRCam filters \citep[F115W, F150W, F200W, F277W, F356W, F444W;][]{Williams2025}. We compute the total sky coverage of all NIRCam imaging considered in this study while carefully accounting for NIRCam’s inter-detector and inter-module gaps as well as the complex survey geometry. The resulting total survey area is 0.28 deg$^2$, corresponding to approximately 1000 arcmin$^2$, the largest to date for a JWST study of $z>3$ quiescent galaxies.

Our dataset combines JWST/NIRCam Cycle 1 pure-parallel observations from the PANORAMIC survey \citep{Williams2025} with archival JWST imaging from the following extragalactic legacy fields:
\begin{itemize}
    \item the EGS field, with data from CEERS (ERS-1345, PI Finkelstein; \citealt{Finkelstein2025}), GO-2279 (PI Naidu), DDT-2750 (PI Arrabal-Haro), and GO-2234 (PI Bañados);
    \item the UDS and COSMOS fields, with data from PRIMER (GO-1837, PI Dunlop), COSMOS-Web (GO-1727, PIs Kartaltepe \& Casey; \citealt{Casey2023}), GO-1810 (PI Belli), GO-1840 (PI Álvarez-Márquez), and DDT-6585 (PI Coulter);
    \item the GOODS fields, with data from JADES (GTO-1180, GTO-1181, PI Eisenstein; GTO-1210, PI Luetzgendorf; \citealt{Eisenstein2023jades}), FRESCO (GO-1895, PI Oesch; \citealt{Oesch2023}), CONGRESS (GO-3577, PIs Egami \& Sun), GO-2079 (PI Finkelstein), JEMS (GO-1963, PIs Williams, Maseda \& Tacchella; \citealt{Williams2023}), JOF (GO-3215, PI Eisenstein; \citealt{Eisenstein2023jof}), GTO-1264 (PI Robledo), GO-4762 (PI Fujimoto), and SAPPHIRES (GO-6434, PI Egami);
    \item the Abell-2744 cluster, with data from UNCOVER (PIs Labbé \& Bezanson; \citealt{Bezanson2024}), GLASS (ERS-1324, PI Treu; \citealt{Treu2022}), DDT-2756 (PI Chen), All the Little Things (GO-3516, PIs Naidu \& Matthee; \citealt{Naidu2024}), and Medium Bands, Mega Science (GO-4111, PI Suess; \citealt{Suess2024});
    \item the JWST North Ecliptic Time-Domain Field, with data from PEARLS (GTO-2738, PIs Windhorst \& Hammel; \citealt{Windhorst2023}).
\end{itemize}
All imaging mosaics were consistently reduced using the software {\sc grizli} \citep{grizli, Valentino2023}, starting from level-2 calibrated data products retrieved from the Mikulski Archive for Space Telescopes (MAST).

The photometric catalog used in this study is from \citet[][to which we refer the reader for details]{Weibel2025uvlf}, constructed from uniformly reduced, PSF-homogenized mosaics following the method described in \citet{Weibel2024} and \citet{Williams2025}. Source detection was performed on inverse-variance weighted stacks of the NIRCam LW images (F277W, F356W and F444W) with the software {\sc SExtractor} \citep{Bertin1996} in its dual-image mode. Fluxes were measured in $r=0.16''$ circular apertures\footnote{The typical half-light size of $z>3$ massive quiescent galaxies is $\lesssim0.5$ kpc \citep{ji2024size}, corresponding to $\lesssim0.1''$. Using a $r=0.16''$ circular aperture -- approximately the FWHM of the F444W PSF -- captures most of the light from these quiescent galaxies while maintaining high S/N photometry. This is essential for obtaining accurate color measurements and, consequently, reliable SED fitting. Missing light is accounted for through our aperture correction based on the Kron flux.} on PSF-matched mosaics of all available JWST and HST bands, and scaled to total using Kron apertures with additional PSF-wing corrections. Empirical PSFs were derived from isolated stars where available; otherwise, model PSFs from {\sc WebbPSF} \citep{Perrin2014} were used after being rotated to match the position angle of observations. 

In Appendix \ref{app:compare_photometry}, we compare our photometry in GOODS-S with the latest public release from the JADES team \citep{Johnson2026,Robertson2026} and find excellent agreement between the two measurements, demonstrating the robustness of our photometric procedure. To avoid mixing catalogs released by different surveys, we adopt our own photometric catalog from \citet{Weibel2025uvlf} across the entire sample, ensuring uniform photometric measurements across all fields.

\section{Sample Selection} \label{sec:selection}

Our parent sample comprises $\sim10^6$ sources after removing artifacts and those with compromised photometry using the \texttt{junk\_flag} and \texttt{hugekron\_flag} from \citet{Weibel2025uvlf}.\footnote{
\citet{Weibel2025uvlf} catalog has a \texttt{use\_phot} flag, which -- in addition to flagging sources with \texttt{junk\_flag} or \texttt{hugekron\_flag} -- flags sources with \texttt{stellar\_flag = True}. We intentionally retained sources with \texttt{stellar\_flag} in the parent sample, as quiescent galaxies at $z > 3$ have typical effective radii $\lesssim 0.1"$ \citep{ji2024size} which is about half the F444W angular resolution.} Running {\sc Prospector} SED fitting (Section~\ref{sec:refine_selection}) to the full catalog is computationally prohibitive\footnote{With our current {\sc Prospector} setup, a single galaxy requires $\sim$10--15 hours.}. We therefore first identified quiescent candidates using established color-selection techniques, followed by {\sc Prospector} fitting and visual inspection, a strategy similar to other high-$z$ quiescent galaxy studies \citep[e.g.,][]{Ji2022,Baker2025}.

Our final $z\sim3$--8 quiescent sample comprises a gold subsample (minimal classification uncertainty) and a more inclusive silver subsample. Their spatial distributions are shown in Figure~\ref{fig:footprints}. Below we describe the selection procedure in detail.

\subsection{Initial selection} \label{sec:initial_selection}

The goal of this initial step is to select quiescent candidates at $z>3$ as broadly and comprehensively as possible. To this end, our selection {\it combines} candidate quiescent galaxies identified using two commonly employed color-selection techniques: (1) NIRCam-based observed colors, and (2) rest-frame colors. Observed colors are effective at isolating broad spectral features (e.g., the Balmer break) with minimal assumptions about galaxy SED templates, but with a fixed filter set, our observed color selection only works within a specific redshift range. In contrast, selecting quiescent galaxies based on rest-frame colors can, in principle, be applied at any redshift. However, deriving rest-frame colors depends on the robustness of SED template fitting (e.g., EAzY; \citealt{Brammer2008}). As each method has its own strengths and limitations, we decided to combine quiescent galaxies selected by either method to construct an initial, comprehensive sample of $z>3$ quiescent candidates.

We select quiescent galaxies at $z\ge3$ using NIRCam observed colors with the selection developed by \citet{Long2024}. Those authors proposed two selection boxes: one with conservative color thresholds that minimize contamination from dusty star-forming galaxies but
may miss dusty quiescent systems, and a more inclusive ``red selection wedge'' (see the Appendix of \citealt{Long2024}). In light of recent discoveries of early massive quiescent galaxies with significant dust reservoirs \citep{Ji2024, Setton2024, Siegel2025}, we chose to use the red selection wedge defined as:
\begin{equation}
    \begin{aligned}
    \rm{(F150W - F277W)} & < 1.5 + 6.25\times\rm{(F277W - F444W)}\\
    \rm{(F150W - F277W)} & > 1.15 - 0.5\times\rm{(F277W - F444W)}\\
    \rm{(F150W - F277W)} & > -0.6 + 2.0\times\rm{(F277W - F444W)}
    \end{aligned}
    \label{equ:red_wedge}
\end{equation}

\begin{figure*}
    \includegraphics[width=0.97\textwidth]{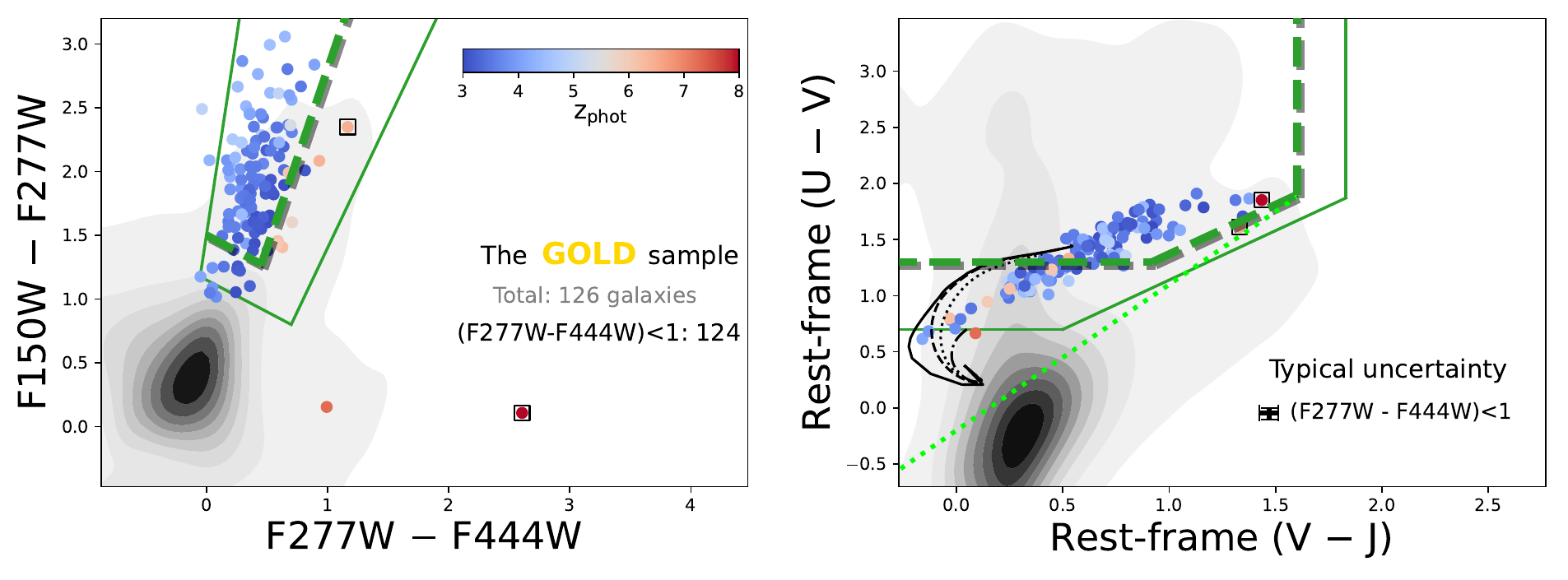}
    \includegraphics[width=0.97\textwidth]{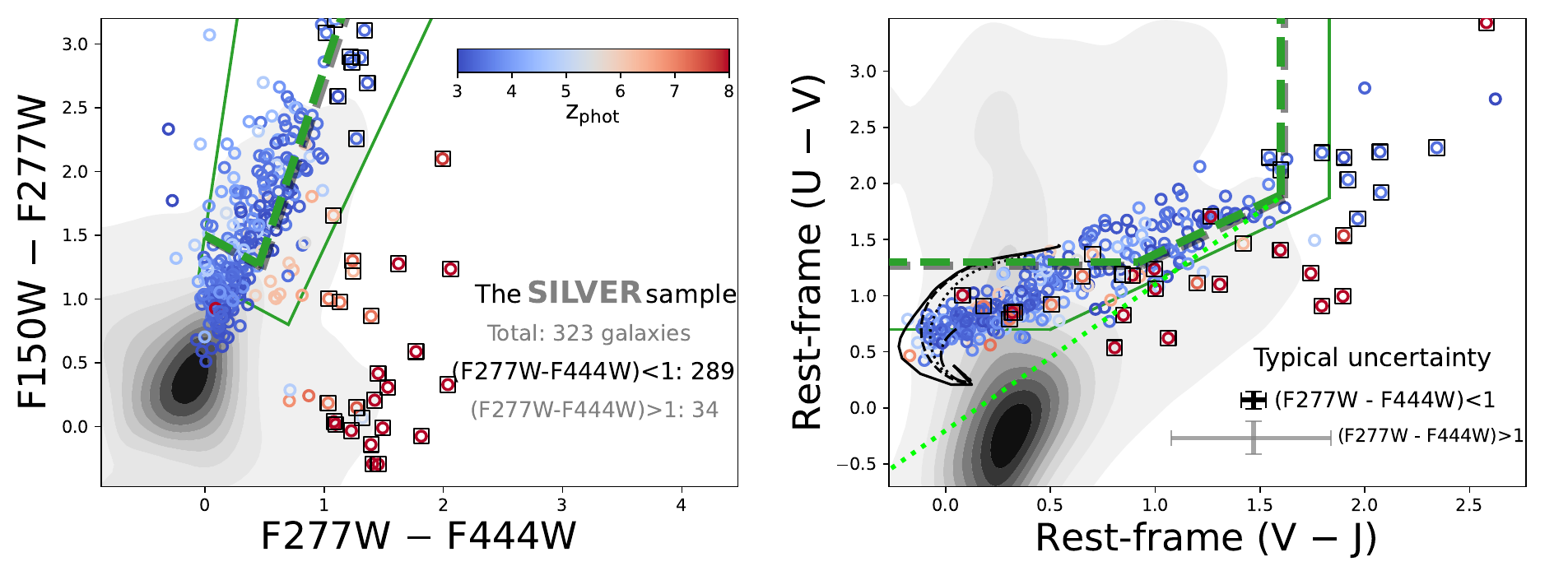}
    \caption{The distribution of $z \sim 3$–8 quiescent galaxies in the observed F150W$-$F277W vs. F277W$-$F444W (left) and the rest-frame $\rm{U-V}$ vs. $\rm{V-J}$ (right) color planes. The upper and lower rows show the gold and silver samples, respectively (Section \ref{sec:refine_selection}). Gray contours show all galaxies in the photometric catalog of \citet{Weibel2025uvlf} with $z_{\rm phot} > 3$ and S/N $>$ 5 in F150W, F277W, and F444W. The green solid lines represent the ``red selection wedge'' \citep{Long2024} and the extended UVJ selection criteria (which are combined for our initial selection). The green dashed lines represent more conservative selection criteria adopted in this study (Section \ref{sec:initial_selection}). The lime dotted line represents the selection from \citet{Baker2025}. Sources removed for red colors, i.e., (F277W$-$F444W) $>$ 1, are marked with open squares. 
    Right panels show typical rest-frame color uncertainties (quantified as the range spanned by the three SED fittings described in Section \ref{sec:refine_selection}). We show these uncertainties separately for silver sample galaxies with F277W$-$F444W $>$ 1 and F277W$-$F444W $<$ 1. The black solid, dashed, dotted, and dash-dot lines in the UVJ plot represent the $0.02 \sim 1$ Gyr evolutionary tracks of the FSPS stellar population models with $\mathrm{E(B-V)} = 0.3$ and a delayed-$\tau$ star formation history, where $\tau = 10, 50, 100,$ and $500$ Myr, respectively.}
    \label{fig:colors}
\end{figure*}

For the rest-frame color selection method, we adopt the rest-frame UVJ  technique \citep{Williams2009}. However, the original UVJ selection box was defined for $z<2$, while quiescent galaxies at $z>3$ are typically younger and thus exhibit bluer colors \citep[e.g.,][]{Belli2019,Valentino2023,Carnall2023,Baker2025}. To account for this, we use an extended UVJ selection that expands the original box to more completely capture $z>3$ quiescent candidates \citep{Valentino2023,AntwiDanso2023,Baker2025b} as follows:
\begin{equation}
    \begin{aligned}
    \rm{(U-V)} & > 0.70\\
    \rm{(V-J)} & < 1.83\\
    \rm{(U-V)} & > 0.88\times \rm{(V-J)}+0.26
    \end{aligned}
    \label{equ:uvj}
\end{equation}
Our criteria closely follow those in \citet{Valentino2023}, who expanded the original UVJ selection box by 0.23 mag. The only difference is the $\rm(U-V)$ threshold: \citet{Valentino2023} adopted $\rm(U-V) > 1$, whereas we further relax it to $\rm(U-V) > 0.70$, consistent with spectroscopically confirmed high-redshift quiescent galaxies with young stellar ages \citep[e.g.,][]{Park2023,Weibel2025}.\footnote{\citet{Weibel2025} did not report rest-frame colors for the RUBIES $z_{\rm spec}=7.29$ quiescent galaxy; we measure $\rm(U-V)=0.9\pm0.1$.}
We also note that \citet{Baker2025} proposed an alternative high-$z$ UVJ box (lime dotted line in Figure \ref{fig:colors}); it allows $\rm(U-V)<0.7$, but all of their spectroscopically confirmed quiescent galaxies have $\rm(U-V)\gtrsim0.7$. Our UVJ selection is therefore also very similar to that of \citet{Baker2025}.

We required S/N $\ge 5$ in the F150W, F277W, and F444W filters to ensure high-quality color measurements and, consequently, reliable SED-fitting results. We imposed a cut of EAzY-derived $z_{\rm phot} \ge 3$ from \citet{Weibel2025uvlf}, and then applied the aforementioned color selection methods to identify quiescent galaxy candidates. In addition, although they constitute only a small fraction ($\approx 2.8\%$), some sources in the \citet{Weibel2025uvlf} catalog have $z_{\rm phot} = -1$, indicating unreliable EAzY fits that may result from the incomplete set of SED templates used in their EAzY configuration. To ensure we did not overlook $z > 3$ quiescent galaxies due to such failed fits, we also included sources with $z_{\rm phot} = -1$ that satisfy the red selection wedge (Equation \ref{equ:red_wedge}).\footnote{Because these sources lack reliable EAzY solutions, the rest-frame UVJ selection could not be applied.}

Our initial selection described above yielded 2633 galaxies in total, of which 2148 and 815 satisfied the NIRCam-based observed-color selection and the rest-frame UVJ selection, respectively. As we will discuss in Section \ref{sec:final_sample}, these two selections are highly consistent with each other at $z > 3$. Nonetheless, combining the two methods allowed us to identify robust quiescent galaxies that would have been missed if only one method had been applied. 

\subsection{Refined selection} \label{sec:refine_selection}

In this section, we use SED fitting and visual inspection to refine our quiescent galaxy selection, exclude contaminants and sources whose emission is likely non-stellar. We used the SED-fitting framework {\sc Prospector} \citep{Johnson2021}, in a similar configuration to \citet{Ji2024jems}. Accurately measuring the SFRs of quiescent galaxies remains challenging \citep{Conroy2013}, as the inferred values are highly sensitive to the assumed star formation history (SFH). We therefore performed SED fitting using three SFH models: a delayed-$\tau$ SFH, and two non-parametric SFHs: one with a continuity prior and one with a bursty continuity prior. The detailed assumptions adopted in our SED analysis and their comparison are presented in Appendix \ref{app:sed_assumption}.

With the SED-fitting results in hand, we refined the initial selection of quiescent galaxies by applying a redshift-dependent cut on sSFR:
\begin{equation}\label{equ:ssfr_thresh}
\rm{sSFR} \le \frac{0.2}{t_H(z)},
\end{equation}
which has been commonly used in recent studies of high-redshift quiescent galaxies \citep[e.g.,][]{Pacifici2016,Carnall2023,Baker2025}. To account for the systematic differences in SFR estimates across the SED models, we define two categories: 
\begin{itemize}
    \item {\bf Gold sample:} A galaxy is assigned to this group if its sSFR satisfies the above threshold in {\it all} three SED fits;
    \item {\bf Silver sample:} the sSFR threshold is satisfied in one/two of the three fits. 
\end{itemize}

We perform a final visual inspection of the image cutouts for all galaxies in the gold and silver samples. This step enabled us to remove objects with clear imaging artifacts that were not flagged by automated pipelines (e.g.,  bright PSF spikes, or regions strongly affected by cosmic-ray hits). In total, we excluded 20 sources through this procedure. Appendix \ref{app:visual_removal} presents detailed notes on the removed sources. While Silver candidates that satisfy the sSFR threshold in only one versus two of the SED fits may have different levels of reliability, the currently available data are too limited to quantify this difference robustly.

\subsubsection{Removal of sources with very red colors} \label{sec:red_removal}

Finally, we excluded sources from the gold and silver samples with very red colors of $\rm F277W-F444W \ge 1$. This decision is motivated by the recent discovery of the so-called little red dots \citep[LRDs; e.g.][]{Kocevski2023,Matthee2024,Greene2024}. Their physical origin remains uncertain, but are likely dominated by a non-stellar power source that can generate strong Balmer breaks, in some cases appearing similar to quiescent galaxies in photometry \citep[e.g.,][]{Williams2024,Labbe2025,degraaff2025cliff,Naidu2026}. Because of this ambiguity, their inclusion in quiescent galaxy samples is potentially problematic. Comprehensively identifying LRDs with photometry alone remains challenging. Photometric selections recover only about half of spectroscopically confirmed LRDs, with the remainder missed due to faint rest-UV fluxes, bluer rest-optical colors, or uncertain redshifts \citep{Hviding2025}. This highlights the incompleteness of purely photometric criteria, though a conservative cut of $\rm F277W-F444W \ge1$ has been demonstrated to remove $\gtrsim60\%$ of spectroscopic LRDs (Section 5.1 of \citealt{Hviding2025}). 

As Figure~\ref{fig:colors} shows, the $\rm F277W-F444W>1$ population (open squares) clusters preferentially at $z_{\rm phot}\sim7$, because the Balmer break redshifts into the $\rm F277W - F444W$ color. At this redshift, using broad band photometry alone, it is difficult to distinguish a rising or flat stellar continuum from strong emission line boosts by e.g., LRDs. Removing  red sources  may also remove  legitimate $z\gtrsim7$ quiescent galaxies. We discuss this issue in detail in Section~\ref{diss:zgt7}.

We note that removing sources with $\rm F277W-F444W>1$ alone does not eliminate all possible LRD-like contaminants from the final sample \citep{Hviding2025}. Cross-matching our final catalog with the \citet{deGraaff2025lrd} compilation of LRDs shows that three such sources remain in the baseline sample, including the Cliff \citep{cliff}. Although this represents only a very small ($<$1\%) fraction of the total, it shows that there is some degeneracy between QGs and LRDs in terms of the photometry. We refer the reader to Weibel et al.\ (2026, in prep.), for a photometric search of sources like the Cliff, and further discussion of the degeneracy with quiescent galaxies. Cross-matching with their sample suggests a small potential contamination of $<5\%$ percent of our quiescent sample and reveals 4 more spectroscopically confirmed LRDs with broad lines. In the analysis that follows, we remove all seven confirmed broad-line LRD sources from our sample, and note that this has no significant impact on our results.

\subsection{Final samples of quiescent galaxies} \label{sec:final_sample}

\begin{figure*}
    \includegraphics[width=1\textwidth]{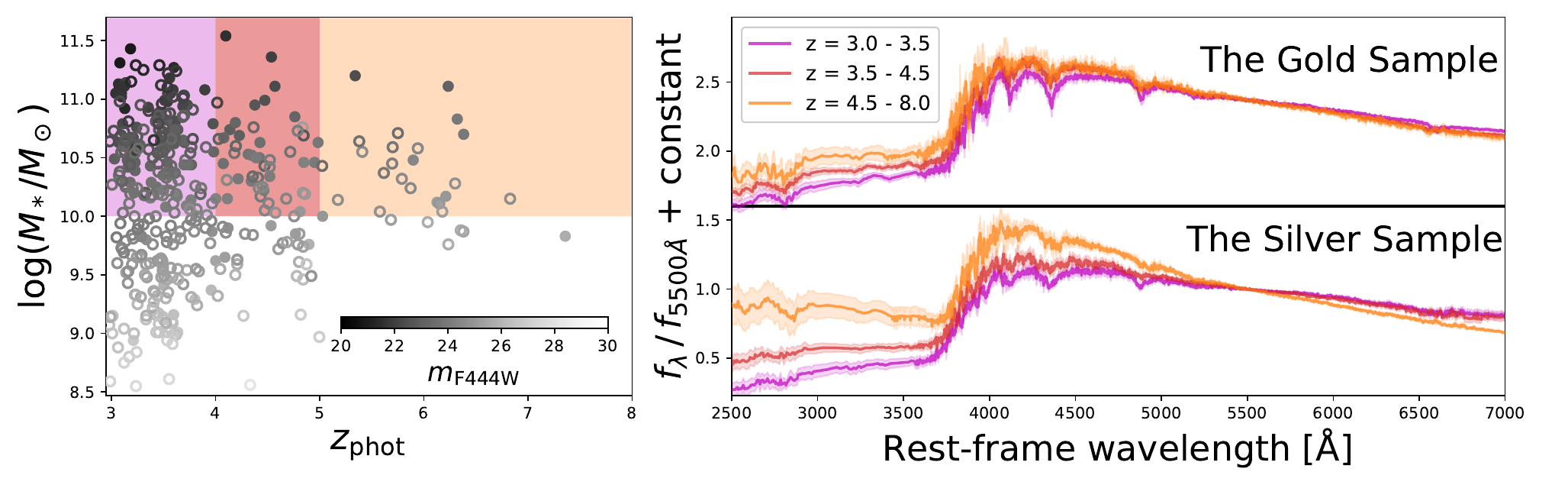}
    \caption{{\bf Left:} $M_*$ vs. $z_{\rm{phot}}$ for our final samples of quiescent galaxies. The gold sample is shown as filled circles, and the silver sample as open circles. This study focuses on massive systems with $M_*>10^{10}M_\odot$. 
    {\bf Right:} Median best-fit SEDs of $>10^{10}M_\odot$ quiescent galaxies in three redshift bins: $3\le z<3.5$ (magenta), $3.5\le z<4.5$ (red), and $z\ge4.5$ (orange). The shaded regions represent the 1-$\sigma$ uncertainties, derived by bootstrapping the galaxy samples in each redshift bin 200 times.}
    \label{fig:median_sed}
\end{figure*}

Our final $z>3$ sample consists of 118 galaxies in the gold sample and 288 galaxies in the silver sample, spanning a stellar mass range of $10^{8.5}$ to $10^{11.5}\,M_\odot$ (Figure \ref{fig:median_sed}). This study focuses on the massive population with $M_*\gtrsim10^{10}M_\odot$, where our sample is highly complete, typically 2 to 5 mag brighter than the corresponding 5 $\sigma$ depths in the shallowest surveys considered here. In this mass regime, we have 101 and 137 galaxies in the gold and silver samples, respectively. The full catalog of 238 massive quiescent galaxy candidates identified in this work can be found at \href{https://zenodo.org/records/22132642}{{\bf this Zenodo link}}.

We cross-matched our final samples with robust, spectroscopically confirmed ones from the literature, finding 32 matches at $z>3$. The details are presented in Appendix \ref{app:crossmatch}; here, we briefly summarize the key conclusions: 29 spectroscopic confirmations are recovered by our final selection (19 in the gold and 10 in the silver samples). This inclusion rate of 29/32 = 90.6\% indicates that our photometric selection is highly complete at the approximately $90\%$ level. Three spectroscopic confirmations are missed by our selection. 2 galaxies (2/32 = 6.25\%) were excluded because their sSFRs are above our threshold (although still $0.3$–$0.5$ dex below the star-forming main sequence). 1 galaxy (3.13\%) was contaminated by bright neighbors causing problematic segmentation in the automatic pipelines. We chose not to add these 3 missed galaxies back into our final samples, as doing so would complicate the selection function and make our results less straightforward to compare to future studies. Including them would not affect any of our conclusions.

The vast majority of our $z>3$ quiescent candidates lie within or close to the quiescent regions in both the observed NIRCam color space and the rest-frame UVJ plane, supporting the robustness of our extended color criteria (green solid lines in Figure \ref{fig:colors}). A more conservative color selection (green dashed lines in Figure \ref{fig:colors}) would miss a substantial fraction of the population. We also find that relying on a single method is incomplete: using only the NIRCam “red selection wedge” would miss 5\% of the gold sample and 26\% of the silver sample, while using UVJ (from Prospector) alone would miss 3\% and 10\%\footnote{These numbers become much higher if the rest-frame UVJ colors are from EAzY, highlighting the large uncertainties in EAzY rest-frame UVJ estimates. A similar finding is reported in \citet{Baker2025b}.}, respectively. Finally, the silver sample is systematically bluer and lies closer to the selection boundaries than the gold sample, consistent with the silver criteria admitting more recently quenched or less securely quiescent systems, whereas the gold sample selects the most robustly quiescent galaxies.

\begin{figure}
    \centering
    \includegraphics[width=1\linewidth]{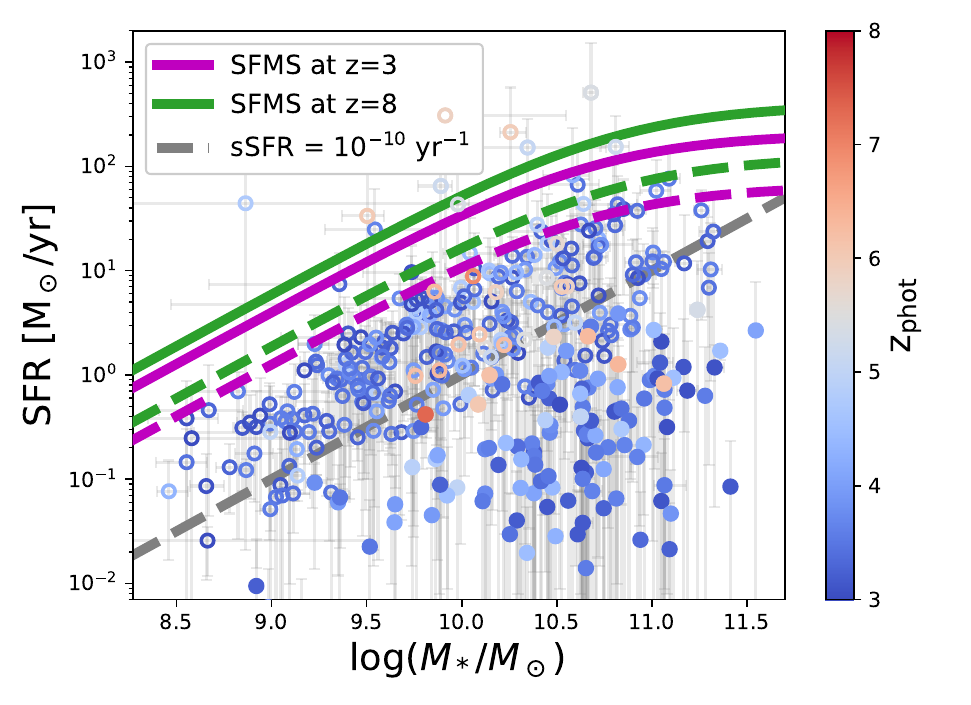}
    \caption{SFR vs.\ $M_*$. Quiescent galaxies in the gold and silver samples are shown as filled and open circles, respectively. Stellar masses and SFRs are plotted as the mean from the three SED fittings described in Section \ref{sec:refine_selection}, with error bars indicating the full range spanned by these fittings. The magenta and green solid lines show the star-forming main sequence at $z=3$ and $z=8$ from \citet{Popesso2023}, while the corresponding dashed lines mark the relation offset by $-0.5$ dex. The grey dashed line marks sSFR $=$ 10$^{-10}$ yr$^{-1}$.}
    \label{fig:sfms}
\end{figure}

In the SFR–$M_*$ plane (Figure \ref{fig:sfms}), our quiescent samples lie well below the star forming main sequence. All galaxies in the gold sample are more than 1 dex below the main sequence, while the silver sample, despite slightly higher inferred sSFRs, also lies systematically below it, with the vast majority offset by more than 0.5 dex (at least 131/137, $>$95\%). Only a very small subset of the full silver sample falls closer to the main sequence (about 10/288, about 3.5\%), indicating that contamination by star forming systems is minimal. Taken together, these trends demonstrate that our photometric selection identifies a robust population of quiescent galaxies at $z>3$, with the gold sample representing the cleanest, most strongly quenched subset.

\begin{figure*}
\centering
    \includegraphics[width=1\textwidth]{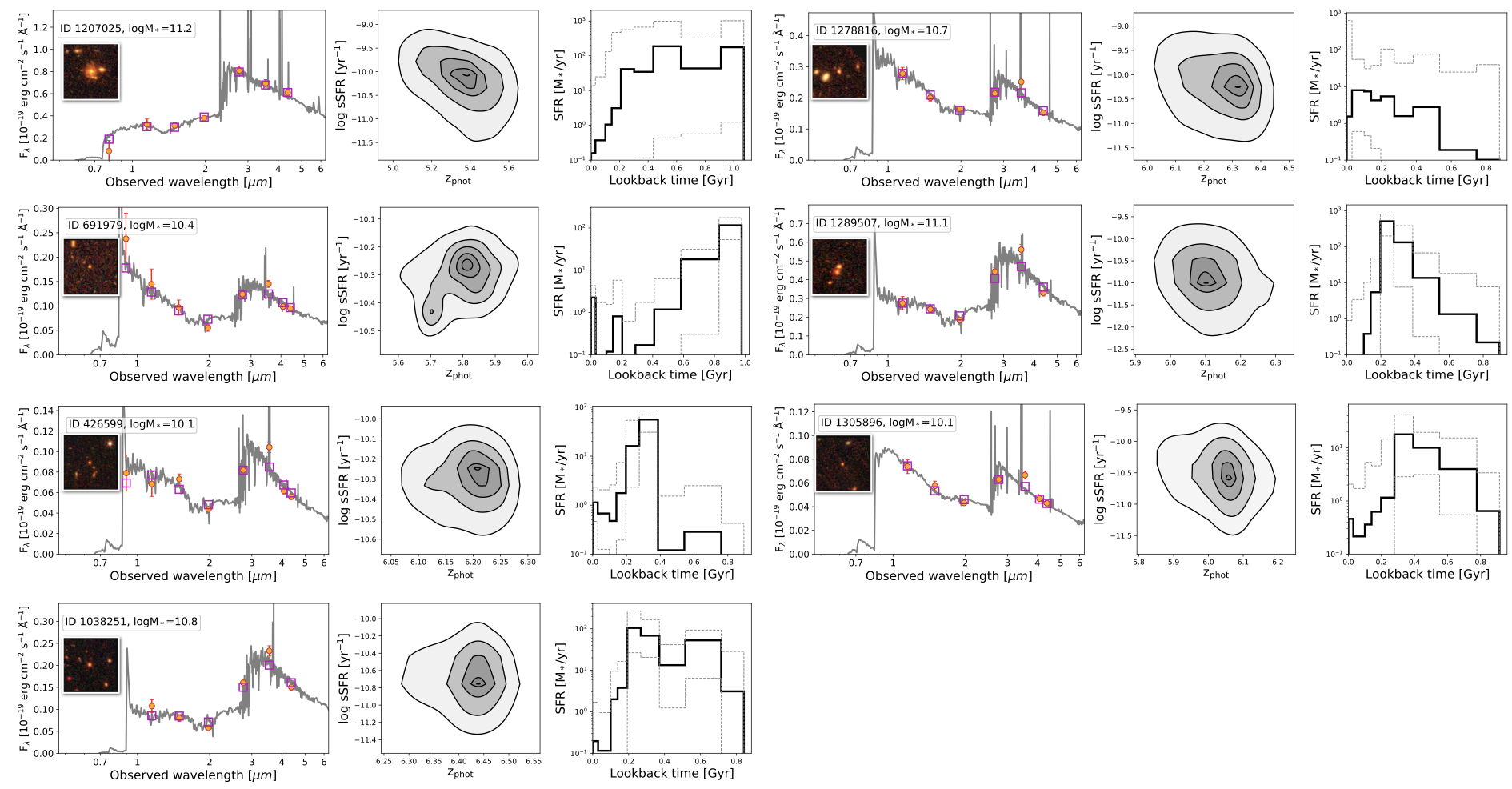}
    \caption{Candidates of massive quiescent galaxies at $z\sim6$ in our gold sample that have not yet been spectroscopically confirmed. For each object, we show the JWST image cutout and observed SED with the best-fit model (left), the posterior constraints on photometric redshift and specific star formation rate (middle), and the inferred star-formation history (right) from the continuity SFH SED fitting. }
    \label{fig:z6}
\end{figure*}

Figure \ref{fig:z6} shows the gold-sample quiescent galaxy candidates at $z\sim 6$ that are still pending spectroscopic confirmation. The photometric coverage provides good sampling of the Balmer break, tracing the SED on both the blue and red sides of rest-frame 4000\AA. The blue-side SEDs resemble those of post-starburst or quiescent galaxies, while at longer wavelengths these objects do not show the unusually red continuum behavior typically associated with LRDs. Together, these features strengthen the interpretation of these systems as high-redshift quiescent candidates, although spectroscopic confirmation is still required.

\section{Number density of massive quiescent galaxies at redshift $z\sim3$-8}\label{sec:n}

Here we present our constraints on the cosmic number density of quiescent galaxies with $M_\ast \geq 10^{10}M_\odot$ at $z \geq 3$, based on 1000 arcmin$^2$ of NIRCam imaging. Our measurements carefully account for both random and systematic uncertainties. 

\subsection{Method} \label{sec:prob_method}
To compute the cosmic number density of photometrically selected quiescent galaxies, many previous studies simply bin quiescent galaxies by redshift and divide by that bin's survey volume \citep[e.g.,][]{Carnall2023,Valentino2023,Alberts2024,Baker2025}. While these studies often account for redshift uncertainties by Monte Carlo resampling of the $z_{\rm phot}$ probability distribution function (PDF), uncertainties in sSFR are typically ignored. Although some earlier works have incorporated the sSFR PDF when defining robust quiescent samples \citep[e.g.,][]{Carnall2020,Carnall2023}, the classification of a galaxy as quiescent has remained deterministic in those studies.

For this study, we combine data from multiple surveys that differ in imaging coverage, depth, and hence the quality of the inferred physical properties. To handle these variations consistently and to propagate measurement uncertainties more robustly, we adopt a probabilistic framework. We make use of the full joint posterior distribution of $z_{\rm phot}$ and sSFR derived from our \textsc{Prospector} SED fitting for each quiescent galaxy to compute their cosmic number density.

For the $i$-th galaxy in the sample, rather than assigning a fixed classification based on a single best-fit value, we evaluate the probability of it being quiescent at redshift $z_k$ as
\begin{equation}
\begin{split}
P_i^{\mathrm{Q}}(z_k) =\ & P_i(z_k) \times \\
& \int_0^{+\infty} \mathcal{S}(\mathrm{sSFR}, z_k) \, P_i(\mathrm{sSFR} \mid z_k) \, d(\mathrm{sSFR}),
\end{split}
\end{equation}
where $P_i(z_k)$ is the redshift posterior probability density, $P_i(\mathrm{sSFR} \mid z_k)$ is the conditional sSFR posterior from our \textsc{Prospector} SED fitting, and $\mathcal{S}(\mathrm{sSFR}, z)$ is the selection function for quiescent galaxies:
\begin{equation}
\mathcal{S}(\mathrm{sSFR}, z) =
\begin{cases}
1 & \text{if } \mathrm{sSFR} < \dfrac{0.2}{t_H(z)}, \\
0 & \text{otherwise.}
\end{cases}
\end{equation}
Summing over the full sample, the expected number of quiescent galaxies in the redshift bin $[z_k, z_k+\Delta z)$ is
\begin{equation}
\begin{split}
N_{\mathrm{Q}}(z_k) = \sum_{i=1}^{N_{\mathrm{gal}}} & \int_{z_k}^{z_k + \Delta z} P_i(z)\times \\
&  \left[ \int_0^{0.2 / t_H(z)} P_i(\mathrm{sSFR} \mid z) \, d(\mathrm{sSFR}) \right] dz,
\end{split}
\end{equation}
and the corresponding number density is
\begin{equation}
n_{\mathrm{Q}}(z_k) = \frac{N_{\mathrm{Q}}(z_k)}{\Delta V(z_k)},
\end{equation}
where $\Delta V(z_k)$ is the comoving volume of the redshift bin, computed from the survey geometry and cosmology. 

This approach naturally incorporates the varying uncertainties of individual measurements, making it well suited for complex samples drawn from multiple surveys. Because galaxy classification above is probabilistic and our selection function for quiescent galaxies depends on both redshift and sSFR, we can exploit the full joint posterior distribution of $z_{\rm phot}$ and sSFR to assess whether some quiescent candidates still have a non-negligible probability of being non-quiescent. This joint posterior depends on the adopted SED assumptions, so we discuss robustness among different SED assumptions in Section \ref{sec:n_results}. 

Moreover, this framework can be applied to larger samples to estimate the number of quiescent galaxies that may be missed by our selection (Section \ref{sec:final_sample}) and, in turn, to evaluate potential systematics in the number-density measurements. We applied this approach to the full set of 2633 galaxies from the initial color-based selection (Section \ref{sec:initial_selection}). We find that the inferred number density of quiescent galaxies would increase by a factor of $\approx1.5$ relative to the gold+silver results reported below in Section \ref{sec:n_results}.

Finally, for the uncertainty in the number-density measurements, we consider three sources of error,
\begin{equation}
\sigma_{\rm{total}}^2 = \sigma_{\rm{measurement}}^2 + \sigma_{\rm{Poisson}}^2 + \sigma_{\rm{CV}}^2
\end{equation}
where $\sigma_{\rm{measurement}}$ is estimated as the 16th–84th percentile range from 200 bootstrap resamplings of our sample; $\sigma_{\rm{Poisson}}$ accounts for Poisson shot noise; and $\sigma_{\rm{CV}}$ accounts for cosmic variance which can be calculated as 
\begin{equation}
    \sigma_{\rm{CV}} = \tilde{\sigma}_{\rm{CV}}\cdot n_Q
\end{equation} 
where $\tilde{\sigma}_{\rm{CV}}$ is fractional. 

To estimate $\tilde{\sigma}_{\rm{CV}}$, we use the code developed by \citet{Jespersen2025}, which is based on the \citet{Moster2011} cosmic-variance calculator, combined with constraints from the {\sc UniverseMachine} simulations \citep{Behroozi2019}. This code computes cosmic variance for 0.5 dex stellar-mass bins in each redshift bin. To obtain the cosmic variance for the stellar-mass bin at $M_\ast \geq 10^{10}M_\odot$, we follow \citet{Valentino2023} and weight the contribution of each 0.5 dex mass bin to the total cosmic variance by the quiescent-galaxy stellar mass function of \citet{Baker2025}.\footnote{The highest-redshift bin of the \citet{Baker2025} stellar mass function (their Table 2) is $z=4$–5. For $z>5$, we adopt the same function as for $z=4$–5.} We compute the cosmic variance for each field separately and then combine them following \citet{Moster2011},
\begin{equation}
\tilde{\sigma}_{\rm{CV}}^2 = \frac{\sum_{i} A_i^2 \tilde{\sigma}_{\rm{CV},i}^2}{\left(\sum_i A_i\right)^2},
\end{equation}
where $A_i$ is the area of the $i$-th field. Although the cosmic-variance error is large for individual fields, as expected for high-mass, high-z galaxies in small fields \citep{Steinhardt2021}, it becomes relatively small when combining our total sky coverage of 1000 arcmin$^2$.

\subsection{Results} \label{sec:n_results}

\begin{table}[t]
    \centering
    \begin{tabular}{|c|c|c|c|c|}
    \hline\hline
         & Redshift & N & Number density  & $\tilde{\sigma}_{\rm{CV}}$ \\
         &  & \# & 10$^{-5}$ Mpc$^{-3}$ & fractional \\
        \hline 
        \multirow{5}{*}{Gold} & (3.0, 3.5) & 36 & $1.81\pm0.53$ & 0.06  \\
                             & (3.5, 4.0) & 37 & $1.52\pm0.48$ & 0.08 \\
                             & (4.0, 5.0) &  21 & $0.54\pm0.39$  & 0.12 \\
                             & (5.0, 6.0) & 2 & $0.09\pm0.07$ & 0.19 \\
                             & (6.0, 8.0) & 5 & $0.07\pm0.06$ & 0.59 \\
        \hline
          & (3.0, 3.5) & 97 & $3.14\pm0.71$ & 0.06 \\
         Gold  & (3.5, 4.0) &  74 & $2.35\pm0.61$  & 0.08 \\
        + & (4.0, 5.0) & 44 & $1.10\pm0.54$  & 0.12 \\
        Silver & (5.0, 6.0) & 14 & $0.20\pm0.11$ & 0.19 \\
                             & (6.0, 8.0) & 9 & $0.10\pm0.07$ & 0.59 \\
    \hline\hline                          
    \end{tabular}
    \caption{Cosmic number density of quiescent galaxies with $M_*\ge 10^{10}M_\odot$ at $z\ge3$. The number density and uncertainty (including the cosmic-variance error) are derived from our probabilistic approach (Section \ref{sec:prob_method}), using the posteriors assuming the delayed-$\tau$ SFH. Using posteriors from alternative SED fits does not change the results (Section \ref{sec:n_results}).}
    \label{tab:n}
\end{table}

\begin{figure*}
\centering
    \includegraphics[width=0.7\textwidth]{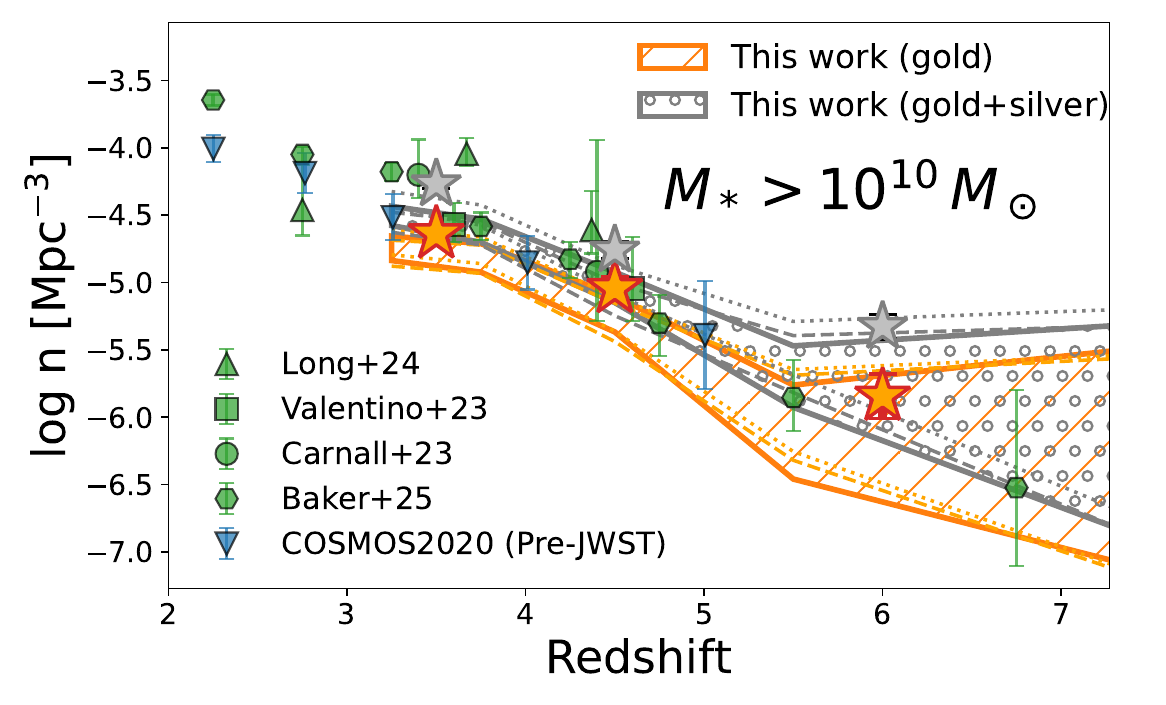}
    \caption{Cosmic number density of $z \geq 3$ quiescent galaxies with $M_\ast \geq 10^{10} M_\odot$ and $\mathrm{sSFR} < 0.2/t_{\rm H}$ (Section~\ref{sec:selection}). Hatched regions show the measurements (1$\sigma$ range) obtained with the probabilistic approach (including the cosmic-variance error) described in Section~\ref{sec:prob_method}, which fully accounts for the posterior distributions of redshift and sSFR. The regions with solid, dashed, and dotted edges correspond to the results based on the delayed-$\tau$, continuity, and bursty continuity SFH fittings, respectively. The large star symbols indicate the measurements from a simple, deterministic approach, where the number density is calculated directly from galaxy samples binned according to their best-fit $z_{\rm phot}$, with only Poisson noise included. For comparison, we show  measurements based on smaller JWST surveys, including \citet[][their Table 5]{Valentino2023}, \citet[][their Table 3, robust subsample]{Carnall2023}, \citet[][their Table 2, NIRCam color selection]{Long2024}, and \citet[][their Table 1]{Baker2025}. We also show pre-JWST measurements from \citet{Weaver2023} based on  COSMOS2020 (other pre-JWST studies are not shown for clarity, but they find generally consistent results, e.g. \citealt{Schreiber2018,Merlin2019,Carnall2020}).}
    \label{fig:ab_obs}
\end{figure*}

Using the probabilistic framework described above, we measured the cosmic number density of massive quiescent galaxies in both the gold and gold+silver samples across the redshift range $z=3$–8. The results are presented in Table \ref{tab:n}, Figure \ref{fig:ab_obs} and \ref{fig:ab_mod}. 

For the gold sample, the number density drops by a factor of $\gtrsim20$ from $z=3$ to $z>6$, decreasing from approximately $2\times10^{-5}$ Mpc$^{-3}$ to $7\times10^{-7}$ Mpc$^{-3}$. At such low abundances (i.e., $\sim10^{-6}$ Mpc$^{-3}$), a sky coverage of $\approx400$ arcmin$^2$ are required to detect a single galaxy over $z=5$–6, underscoring the necessity of wide-area surveys to robustly constrain the number density of the earliest massive quiescent galaxy populations. 

As shown in Figure \ref{fig:ab_obs}, the number densities derived from the probabilistic approach, using posteriors from different SFH fittings (gold regions with different linestyle edges), are highly consistent with each other for galaxies in the gold sample. They are also in excellent agreement with those obtained from a much simpler deterministic approach (gold star symbols), indicating that
\begin{equation}
\int_0^{0.2 / t_H(z)} P_i(\mathrm{sSFR} \mid z)\,d(\mathrm{sSFR}) \approx 1 ,
\end{equation}
i.e., nearly all of the PDF mass lies in the quiescent solution, regardless of the SED assumptions. This demonstrates the very high purity of the gold sample, with galaxies in it having very high confidence of being genuinely quiescent (also see Section \ref{sec:final_sample} and Figure \ref{fig:sfms}).

When the silver sample is included in our analysis, we observe an increase in the number density of quiescent galaxies (gray hatched regions in Figure \ref{fig:ab_obs}). The magnitude of this increase depends on the SED assumptions. Relative to the gold sample alone, the number density of the combined gold and silver samples increases by a factor of $\approx1.6$ when using $z_{\rm phot}$–sSFR posteriors from the delayed-$\tau$ and continuity SFH fittings, and by a factor of $\approx2$ when using posteriors from the bursty continuity fitting, thereby highlighting the systematic biases introduced by different SED assumptions. 
Importantly, the increase persists across all SFH priors. This is consistent with many silver objects carrying high posterior probability of being quiescent under our SED assumptions. Together with their broadly quiescent-like colors (Section \ref{sec:final_sample}), this is indicative that the silver sample contains a real quiescent component, even if its purity cannot be established without independent spectroscopic confirmations.

Our measurements are in broad agreement with previous studies before the launch of JWST. Figure \ref{fig:ab_obs} shows that our new constraints are consistent with the results of \citet[][blue triangles]{Weaver2023} based on the pre-JWST COSMOS dataset. This is likely due to the fact that massive quiescent galaxies at $z>3$ are bright (typically $23-24$ mag in F444W), and thus they were detected and easily characterized by Spitzer/IRAC at $3-4\,\mu$m \citep{Weaver2022}.

\begin{figure}
    \centering
    \includegraphics[width=0.97\linewidth]{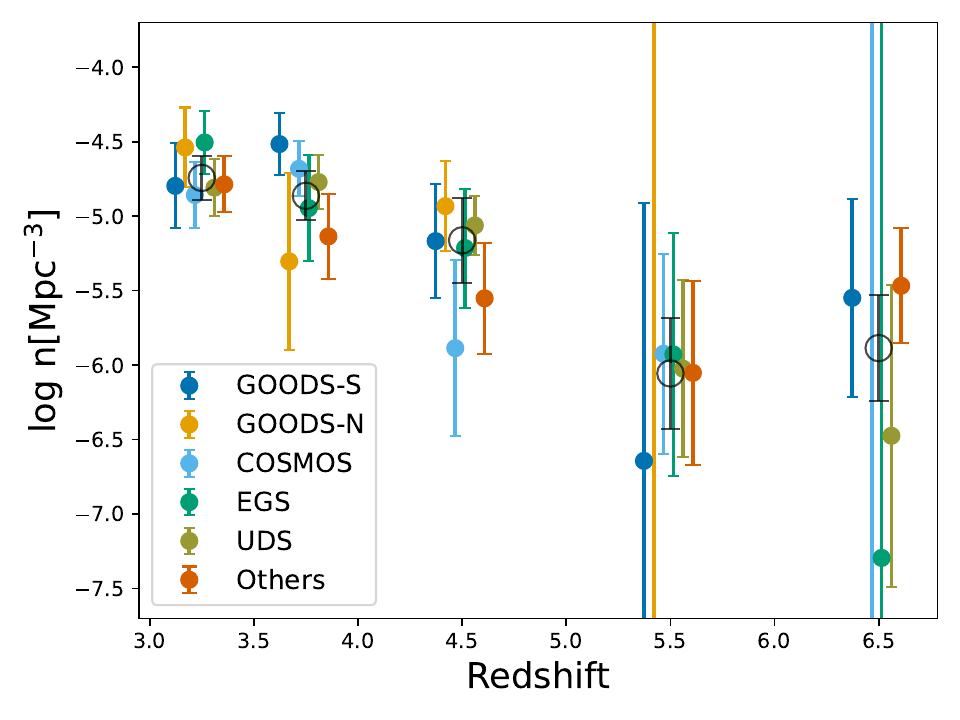}
    \caption{Similar to Figure \ref{fig:ab_obs}, but showing the number density of quiescent galaxies in the gold sample measured in individual fields. The black open circles indicate the results obtained by combining all fields together. The redshift values for each field have been slightly displaced for clarity.}
    \label{fig:ab_fields}
\end{figure}

Our constraints are consistent with earlier JWST studies based on much smaller NIRCam sky coverage (green symbols in Figure \ref{fig:ab_obs}). We also measure the number density of gold-sample quiescent galaxies separately in each subfield. Figure \ref{fig:ab_fields} shows evidence for an enhanced abundance of quiescent systems in EGS at $z=3$–3.5, consistent with \citet{Shuowen2024}. In the $z=3.5$–4 bin, we find an excess in GOODS-S, in agreement with \citet[][their Figure 6]{Baker2025gds}, who reported a similar over-abundance using the photometrically selected quiescent sample from \citet{Ji2024}. At $z=4$–5, we find tentative evidence for an excess of quiescent galaxies in GOODS-N and UDS. Although direct number-density measurements are not yet available for these two fields, we note the recent discovery of two spectroscopically confirmed massive quiescent galaxies at $z\sim4.6$ in PRIMER/UDS from the EXCELS survey \citep{Carnall2024excel}. The two galaxies lie within $<1$ Mpc of each other, hinting at the presence of large-scale structure in UDS at this redshift \citep{Carnall2024excel, Jespersen2025b}.

\subsection{Comparing with model predictions}\label {sec:ab_comp_mod}

A central goal of this paper is to place our measured number densities of massive quiescent galaxies at $z>3$ in the broader context of galaxy-formation models. The redshift evolution of the abundance provides a direct, population-level constraint on the mechanisms that assemble massive systems and then rapidly shut down their star formation.

Throughout this section we implicitly assume that the stellar masses and star-formation rates inferred from the observations are directly comparable to the corresponding quantities reported by models. In practice, however, these quantities are not guaranteed to be defined or measured in an identical way: the models provide ``intrinsic'' physical properties, whereas observational estimates depend on SED-fitting assumptions and can suffer systematic offsets relative to the underlying values. Quantifying the impact of such systematics is non-trivial, and would ideally require end-to-end synthetic observations built from the models and analyzed with the same pipelines used for the data. Such synthetic datasets are not yet available for the full set of models considered below, and producing them is beyond the scope of this work. With this caveat in mind, we present comparisons to different sets of models in which quenching emerges from a range of (often unresolved, sub-grid) prescriptions for gas accretion, star formation, and feedback.

\begin{figure*}
\centering
    \includegraphics[width=0.47\textwidth]{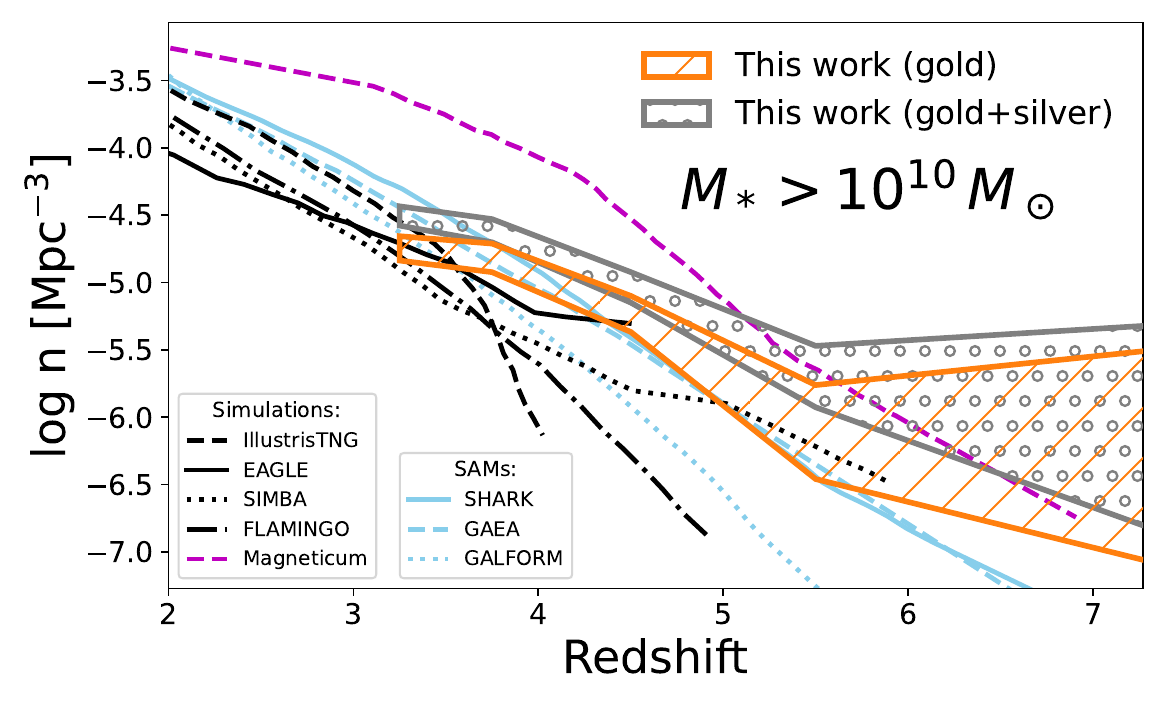}
    \includegraphics[width=0.47\textwidth]{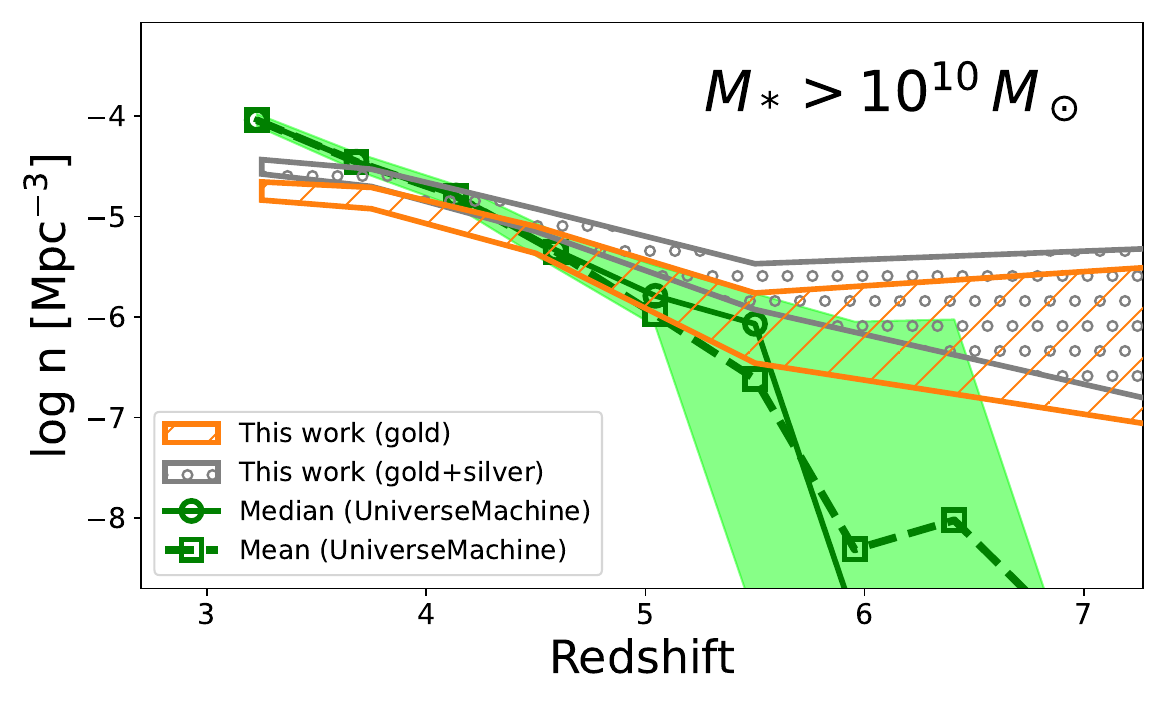}
    \caption{Comparison of the cosmic number density of quiescent galaxies with predictions from simulations and SAMs (left) and from the UniverseMachine empirical framework (right). All models adopt the same stellar mass cut ($\geq10^{10}M_\odot$) and sSFR threshold ($<0.2/t_{\rm H}$) as the observations. Simulations include {\sc IllustrisTNG} \citep{Pillepich2018}, {\sc Eagle} \citep{Schaye2015}, {\sc Simba} \citep{Dave2019}, {\sc Flamingo} \citep{Schaye2023}, and {\sc Magneticum} \citep{Remus2025,Kimmig2025}; SAMs include {\sc Shark} \citep{Lagos2024}, {\sc Gaea} \citep{DeLucia2024}, and {\sc Galform} \citep{Lacey2016}. Simulation and SAM curves are taken from \citet{Baker2025} and \citet{Lagos2025}. UniverseMachine predictions are described in Section \ref{sec:um_ab}; the light green band indicates the 1$\sigma$ (16th to 84th percentile) range, and the y axis limits differ between panels to highlight the predictions at $z>6$ from UniverseMachine. To avoid overcrowding, we show only measurements from the probabilistic approach based on delayed $\tau$ SED fitting posteriors (shaded regions with solid edges in Figure \ref{fig:ab_obs}).
    }
    \label{fig:ab_mod}
\end{figure*}

A growing number of studies have found substantial tension between theoretical predictions and the observed number densities of massive quiescent galaxies, particularly at $z\gtrsim3$ \citep[e.g.,][]{Valentino2023,Alberts2024,Carnall2022,Baker2025b}. We find a similar disagreement when comparing our measurements to a broad suite of state-of-the-art semi-analytic models (SAMs) and cosmological hydrodynamical simulations (Figure~\ref{fig:ab_mod}; models listed therein).

Overall, the hydrodynamical simulations span a wide range of predictions, but several widely used frameworks underpredict the abundance of massive quiescent galaxies at $z\gtrsim4$ by $\gtrsim 1$ dex, with discrepancies that typically grow toward earlier epochs. The SAMs shown in Figure \ref{fig:ab_mod} generally predict higher abundances than the hydrodynamical simulations at $z\gtrsim3$, and some (i.e., SHARK and GAEA, light-blue solid and dashed lines in Figure \ref{fig:ab_mod}) approach the lower envelope of our gold-sample constraints out to $z\sim5$; nonetheless, the SAMs considered here still underpredict the abundance of our gold$+$silver sample at $z\gtrsim4$. We return to this point in Section~\ref{diss:implication}, where we discuss plausible explanations for why some SAMs appear to better reproduce the abundance of quiescent galaxies at $z>3$ and the implications for the physical drivers of early quenching.

\subsubsection{Empirical models: UniverseMachine}
\label{sec:um_ab}

While the SAMs/simulations above are often used as tests of specific physical quenching pathways in galaxy-formation models, empirical models anchored to key observed correlations at lower redshifts (e.g., stellar mass functions and the SFR--$M_*$ relation) provide a deliberately ``physics-light'' but equally stringent baseline for the expected buildup of quiescent populations. In this case, consistency or inconsistency offers a direct diagnostic of whether relations calibrated at lower redshift can be reliably extrapolated into the $z>3$ regime: agreement would suggest that the demographic link between mass growth and quenching inferred at later times already captures the essential ingredients of quenching at early epochs, whereas tension would imply that the high-$z$ Universe requires additional evolution in quenching efficiency, timescales, or triggering channels beyond what is encoded by lower-$z$ calibrations.

We compare our measurements to the {\sc UniverseMachine} empirical framework \citep{Behroozi2019}, which assigns star-formation histories to dark-matter halos in a way that is statistically constrained by a wide range of observational demographic measurements across cosmic time. We generate mock realizations matched to the exact survey geometry of each pointing shown in Figure~\ref{fig:footprints}. To better sample the range of large-scale environments, we generate 16 light cones for each pointing. In each Monte Carlo realization, we randomly select one light cone per pointing, combine them to form a mock survey with the same total footprint as the data, apply the same quiescent selection (Section~\ref{sec:selection}), and compute the number density of massive quiescent galaxies at $z>3$. Repeating this procedure $10^{4}$ times, we measure the median and mean predicted number densities, and the $1\sigma$ scatter (taken as the 16th--84th percentile range) among realizations. The results are shown in the right panel of Figure~\ref{fig:ab_mod}. 

{\sc UniverseMachine} broadly reproduces the rapid decline of the quiescent-galaxy number density over $z=3$--5. The median and mean predictions lie close to our observational constraints over $z\simeq3.5$--5, while at $z\simeq3.0$--3.5 the model tends to predict a higher abundance of massive quiescent systems than we infer. This is notable because {\sc UniverseMachine} includes a quenched-fraction constraint at $z\sim3$--4 based on the UltraVISTA/COSMOS measurements of \citet{Muzzin2013}, which are dominated by galaxies more massive than those in our sample, with $M_*\gtrsim10^{11}M_\odot$. This difference is unlikely to be driven by cosmic variance: UltraVISTA covers $\simeq1.62{\rm deg}^2$, about six times the total area considered here, while our sample is distributed over 34 independent sightlines. The expected cosmic variance of our combined sample, $\sim0.06$--0.08 at $z\sim3$--4 (Table \ref{tab:n}), is comparable to the $\sim0.08$--0.10 estimated for UltraVISTA \citep{Muzzin2013}. Moreover, our measurements are in good agreement with the more recent COSMOS2020 measurements of \citet{Weaver2023}. Since both UltraVISTA and COSMOS2020 probe galaxies in the COSMOS field, the difference in their inferred quiescent abundances highlights the sensitivity of high-redshift quiescent-galaxy measurements to photometric redshifts and SED-based stellar masses and SFRs. The tension at $z\simeq3.0$--3.5 may therefore largely reflect uncertainties in the earlier quenched-fraction constraints used to calibrate {\sc UniverseMachine}.

At higher redshifts, the {\sc UniverseMachine} predictions drop rapidly: the median and mean abundances decline sharply beyond $z\gtrsim5$ and fall well below our inferred number densities, while the predicted $1\sigma$ scatter increases substantially at $z>5$.  We also observe that the mean prediction exceeds the median at $z\gtrsim6$, indicating a skewed distribution of the {\sc UniverseMachine} predictions, in which most realizations contain very few (or no) such systems, with the mean boosted by a small tail of rare light cones that include one or more massive quiescent galaxies. Because {\sc UniverseMachine} is an empirically calibrated model rather than a fully physics-forward treatment of galaxy quenching, this comparison offers a particularly direct test of extrapolatability: the growing tension at $z\gtrsim5$ implies that the onset and/or efficiency of quenching at the earliest times, and/or the rapid mass growth required to populate the massive end by these epochs, is not fully captured by lower-$z$-anchored empirical prescriptions. In other words, within the current {\sc UniverseMachine} extrapolation, massive quiescent galaxies at $z>5$ are expected to be exceedingly rare, and matching our measured abundances would require substantially earlier and/or more efficient quenching (and/or its maintenance) than is implied by relations calibrated at lower redshift.

\section{Cosmic variance of massive quiescent galaxies at redshift $z\sim3$--8}
\label{sec:cv}

Beyond the mean number-density evolution, the field-to-field variance in galaxy number density carries additional information about their connection to dark matter halo properties and large-scale structure. In pencil-beam surveys, these fluctuations are often dominated by cosmic variance, i.e., the clustering-driven excess (relative to Poisson) variance obtained by integrating the power spectrum weighted by the survey window function \citep{Newman2002,Somerville2004,Trenti2008,Moster2011, Steinhardt2021}. With the multi-pointing geometry and sample size available here, we can, for the first time in the $z>3$ regime, empirically quantify the spatial clustering of massive quiescent galaxies across independent sightlines. Translating the measured cosmic variance into a unique characteristic halo mass is non-trivial, since the derived clustering amplitude relies on non-linear modes. Nevertheless, the observed field-to-field scatter nevertheless provides a direct, model-independent constraint on the clustering strength (i.e., effective bias) of these systems \citep[e.g.,][]{Robertson2010}.

\subsection{Method}

\subsubsection{Cosmic variance estimations}
\label{sec:cv_method}

We measure the fractional cosmic variance ($\sigma_{\rm CV}$) in the  galaxy number counts per NIRCam pointing ($N$) using two complementary estimators. First, we perform a bootstrap variance-decomposition measurement following \citet{Moster2011}. This method does not require assuming a functional form for the galaxy count distribution, but it can be biased high if a small number of outliers inflate the variance. The observed field-to-field variance in galaxy counts receives contributions from Poisson counting noise and from cosmic variance:
\begin{equation}
\sigma_{\rm CV}^2 \;=\; \frac{\langle N^2\rangle-\langle N\rangle^2-\langle N\rangle}{\langle N\rangle^2},
\label{eq:sigcv}
\end{equation}
where $\langle N\rangle$ is the mean number of galaxies per pointing and the Poisson contribution to the variance is $\sigma_{\rm P}^2=\langle N\rangle$.
Uncertainties on $\sigma_{\rm CV}$ are estimated from 10,000 bootstrap resamples of the independent pointings (with replacement), re-computing Equation \ref{eq:sigcv} for each realization, and summarizing the resulting $\sigma_{\rm CV}$ distribution using the median and 16th/84th percentiles. In each bootstrap realization, the estimator above is applied to the independent PANORAMIC pointings shown in Figure~\ref{fig:footprints}, except for where two pointings are separated by less than $1^\circ$; in these cases, we randomly select one pointing from each region to preserve spatial independence.
For the legacy fields (GOODS-S, GOODS-N, Abell~2744, EGS, UDS, and COSMOS), whose footprints extend beyond a single NIRCam pointing, we instead randomly select a single NIRCam pointing within each field (random center and position angle), while requiring that the full footprint of the chosen pointing lies within the legacy-field coverage (the same procedure was used in \citealt{Weibel2025b}).

A more principled approach for quantifying the cosmic variance is to directly fit the galaxy number count distribution. We thus also estimate $\sigma_{\rm CV}$ using a Bayesian distribution-fitting approach that models the full field-to-field count distribution, thereby using all information in the observed distribution. Following \citet{Jespersen2025}, we model the underlying positive-definite distribution of counts with a Gamma functional form, which allows the mean and variance to vary independently and recovers the Poisson-like limit as ${\rm Var}(N)\rightarrow \langle N\rangle$. As noted by \citet{Jespersen2025}, the choice of functional form (Gamma versus Negative Binomial) is empirically negligible. To reduce sensitivity to rare, highly overdense pointings in the skewed count distribution, we define the likelihood using a Negative Binomial form evaluated on the binned number counts, which effectively down-weights extreme overdensities compared to a pure Poisson likelihood \citep{Hogg2010}. We parameterize the model in terms of the mean $\mu\equiv\langle N\rangle$ and $\sigma_{\rm CV}$, using the variance relation implied by Equation \ref{eq:sigcv}:
\begin{equation}
{\rm Var}(N) \;=\; \mu \;+\; \mu^2\,\sigma_{\rm CV}^2.
\label{eq:var_mu_sigcv}
\end{equation}
We fit the data and sample the posterior using the MCMC sampler \texttt{emcee} \citep{ForemanMackey2013}. Following \citet{Weibel2025b}, we adopt a Jeffreys prior on the total variance, $p(\sigma)\propto \sigma^{-1}$, which favors small variance unless a larger variance is strongly supported by the data \citep{Jeffreys1946}. Posterior constraints on $\mu$ and $\sigma_{\rm CV}$ are reported using the median and 16th/84th percentiles of the marginalized posteriors.

\subsubsection{Constructing abundance-matched samples from UniverseMachine}
\label{sec:cv_method_UM}

To compare our cosmic-variance measurements with theoretical expectations, we construct mock quiescent-galaxy samples from the {\sc UniverseMachine} catalogs. As discussed in Section~\ref{sec:um_ab}, applying the same quiescent-galaxy selection used for our observations yields a {\sc UniverseMachine} quiescent number density at $z>3$ that is inconsistent with the data. Because cosmic variance is a second-order statistic that depends on the tracer population, a comparison of field-to-field fluctuations is not readily interpretable when the underlying mean abundance (the first-order statistic) already differs between the data and the mocks.

We therefore construct abundance-matched {\sc UniverseMachine} samples by enforcing agreement in the mean number density. In each redshift bin, we rank {\sc UniverseMachine} galaxies with $M_*\ge10^{10}M_\odot$ by their sSFR and select the $N$ galaxies with the lowest sSFR, where $N$ is chosen such that the resulting {\sc UniverseMachine} number density matches the observed quiescent-galaxy abundance in the same redshift interval (Table~\ref{tab:n}). This procedure is equivalent to applying a redshift-dependent sSFR threshold, yielding samples with matched first-order statistics while largely preserving {\sc UniverseMachine}'s predicted clustering for an abundance-matched population. The implied $\log(\mathrm{sSFR/yr^{-1}})$ thresholds are $-11.5$, $-9.5$, $-9.4$, $-9.3$, and $-8.9$ from low to high redshift, respectively. Relative to the $0.2/t_{\rm H}$ criterion, these thresholds are lower in the lower-redshift bins (e.g., $\sim10^{-10}\ \mathrm{yr^{-1}}$ at $z=3.5$) and larger in the highest-redshift bin (e.g., $\sim10^{-9.5}\ \mathrm{yr^{-1}}$ at $z=7$). Similar to Section \ref{sec:um_ab}, in total we generate $10^{4}$ realizations of the mock observations and use them to derive the median and 16th/84th percentiles of the resulting $N$ distribution.

Our choice of an sSFR-ranked abundance-matching scheme is motivated by two considerations. First, it is robust to potential systematic offsets in the absolute sSFR scale of {\sc UniverseMachine} at high redshift. {\sc UniverseMachine} assigns SFRs empirically by calibrating a flexible model to reproduce a broad set of ensemble observables over the redshift range where constraints exist \citep{Behroozi2019}. However, the quenched component of the SFR distribution is itself poorly constrained observationally, so {\sc UniverseMachine} fixes the lognormal parameters of the quiescent population to values based on $z\sim0$ data. As a result, the absolute sSFR scale of quiescent galaxies in the model, particularly at high redshift, should be regarded as uncertain. A rank-based selection is therefore less affected by monotonic offsets and avoids selecting fundamentally different populations simply because the SFR scale differs between model and data. Second and more importantly for our purposes, since cosmic variance depends on the tracer bias, matching the comoving number density provides a practical way to define samples that are more comparable in their typical host-halo population than a hard sSFR cut when the baseline abundances disagree. Finally, observational quiescent selections (e.g., sSFR-based) effectively isolate the low-sSFR tail at fixed stellar mass; our procedure provides a model-internal analogue of this selection while preserving {\sc UniverseMachine}'s predicted clustering for an abundance-matched population.

\subsection{Results}
\label{sec:cv_results}

\begin{figure}  
    \centering
    \includegraphics[width=1\linewidth]{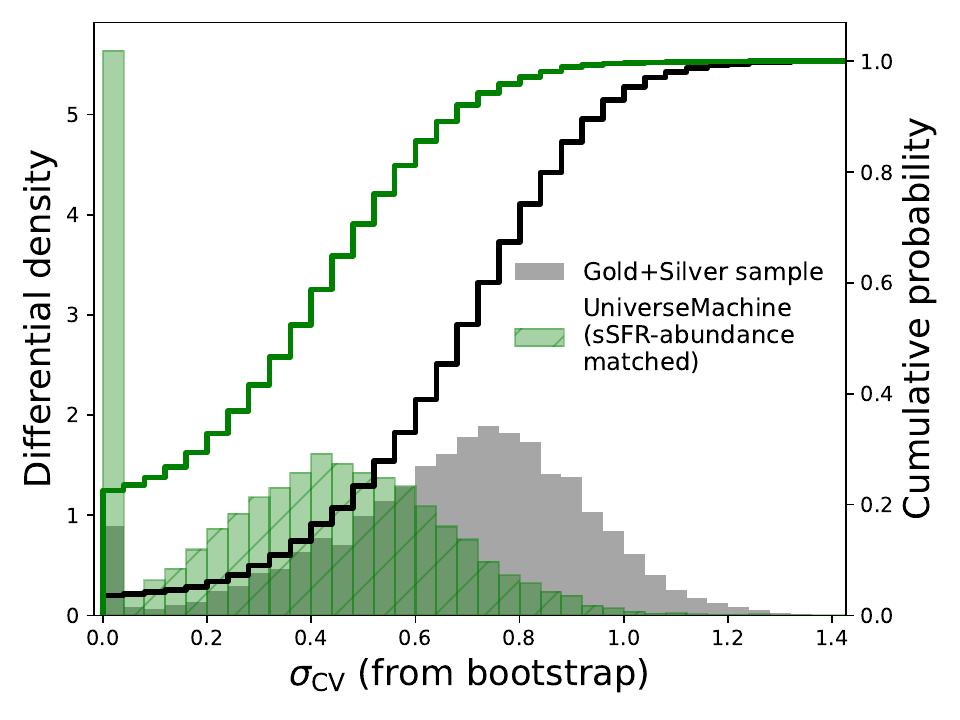}
    \caption{Comparison of bootstrap-inferred cosmic variance (Equation \ref{eq:sigcv}, on the scale of a single NIRCam pointing), $\sigma_{\rm CV}$, between the observed gold+silver quiescent sample and the sSFR-abundance-matched {\sc UniverseMachine} mocks using the same number of sightlines. The filled histograms show the differential distributions of $\sigma_{\rm CV}$, while the solid curves show the corresponding cumulative probabilities (right axis). The observations exhibit systematically larger field-to-field fluctuations than the mocks in two complementary senses: (i) the observed distribution is shifted to higher values, with a larger median $\sigma_{\rm CV}$ than {\sc UniverseMachine}, and (ii) the {\sc UniverseMachine} distribution includes substantial probability mass at $\sigma_{\rm CV}\simeq 0$, with $\sim25\%$ of bootstrap realizations returning $\sigma_{\rm CV}=0$. See Section~\ref{sec:cv_results} for discussion of the interpretation of the low-$\sigma_{\rm CV}$ tail in the {\sc UniverseMachine} mocks.}
    \label{fig:cv_com_boots}
\end{figure}

\begin{figure*} 
    \centering
    \includegraphics[width=0.927\textwidth]{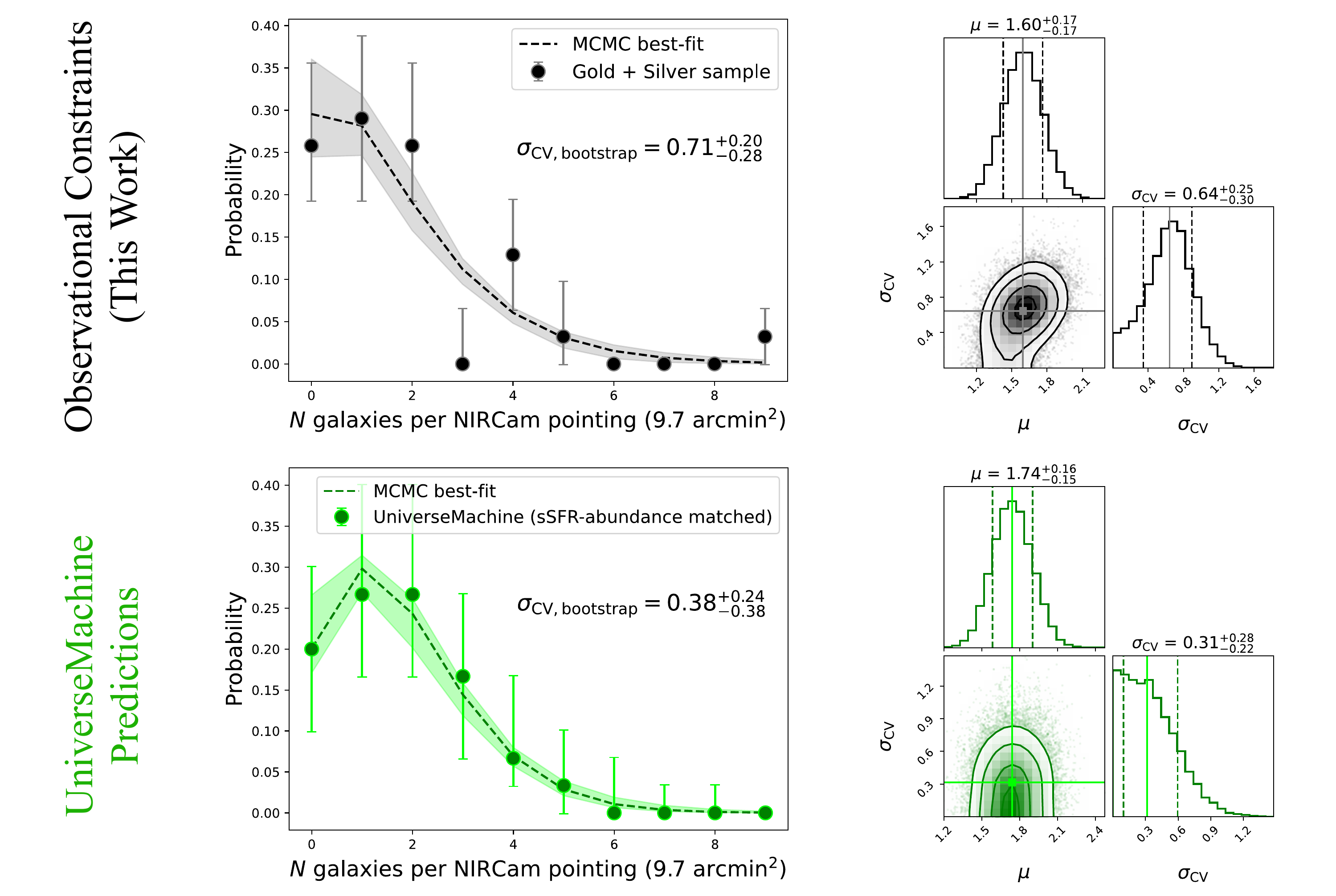}
    \caption{MCMC-inferred cosmic variance. {\bf Top:} Observational constraints for our gold$+$silver sample of massive quiescent galaxies at $z>3$. Left panel shows the probability distribution of the number of galaxies per pointing, $N$ (black points with uncertainties from bootstrap resampling). The dashed curve indicates the MCMC best-fit model (see Section \ref{sec:cv_method}), with the shaded band showing the posterior predictive 1-$\sigma$ range. Right panel shows the corner plot and marginalized posteriors. {\bf Bottom:} Same as the top row, but for the sSFR-abundance-matched {\sc UniverseMachine} mock catalog (see Section \ref{sec:cv_method_UM}), yielding a lower $\sigma_{\rm CV}$ relative to our observational constraints, consistent with our conclusions based on bootstrap-inferred cosmic variance (Figure \ref{fig:cv_com_boots}).}
    \label{fig:cv_com}
\end{figure*}

We first report the cosmic-variance constraints for our quiescent-galaxy samples. The gold sample alone lacks sufficient statistics for a robust cosmic-variance measurement (see Appendix~\ref{app:gold}), so in what follows we focus on the combined gold+silver sample. For the gold+silver sample we measure $\sigma_{\rm CV}=0.71^{+0.20}_{-0.28}$ using the bootstrap variance-decomposition estimator (Figure \ref{fig:cv_com_boots}) and $\sigma_{\rm CV}=0.65^{+0.25}_{-0.30}$ from the Bayesian MCMC distribution-fitting method (the top row of Figure~\ref{fig:cv_com}). We note that the agreement between these two approaches does not generally imply that the two estimators are equally robust: the bootstrap variance-decomposition estimator is more sensitive to outlier fields, and our sample likely does not contain an unusually extreme field.

For the sSFR-abundance-matched {\sc UniverseMachine} mock sample, we infer systematically smaller cosmic variance. From bootstrapping (Figure \ref{fig:cv_com_boots}), we find $\sigma_{\rm CV}=0.38^{+0.24}_{-0.38}$, where the $1\sigma$ lower bound is consistent with zero. Similarly, as shown in the bottom row of Figure~\ref{fig:cv_com}, the MCMC fit yields $\sigma_{\rm CV}=0.31^{+0.28}_{-0.22}$. The posterior for $\sigma_{\rm CV}$ is strongly asymmetric, with its maximum toward very small values (bottom-right panel). It is important to emphasize that this concentration of posterior mass near $\sigma_{\rm CV}=0$ does not imply that the {\sc UniverseMachine} galaxies are not clustered. Rather, it indicates that, \textit{for the same finite number of sightlines (34) as in our observational sample}, a non-negligible fraction of {\sc UniverseMachine} realizations do not produce a statistically measurable excess variance above Poisson noise (Appendix \ref{app:cv_snr} provides a detailed discussion on the detectability of $\sigma_{\rm CV}$). In the limit of an arbitrarily large number of sightlines, the intrinsic cosmic variance of the sSFR-matched {\sc UniverseMachine} sample is $\sigma_{\rm CV}\sim0.43$, which is lower than our current observational constraint at mild significance.

Before interpreting this difference, we first ask whether the observed excess $\sigma_{\rm CV}$ could simply arise from random observational uncertainty. Random redshift or quiescent-classification errors could  broaden the observed field-to-field count distribution, thereby increasing the measured $\sigma_{\rm CV}$ above the intrinsic value. To assess the scale of this effect, we make a simple order-of-magnitude estimate. For the combined massive gold+silver sample used in our main cosmic-variance analysis, we have 238 galaxies across 34 independent sightlines, corresponding to a mean of $\mu \approx 238/34 = 7$ objects per field, while the measured excess variance is $\sigma_{\rm CV}\approx 0.7 \pm 0.3$. If each potentially ambiguous source has an independent probability $p$ of being scattered into or out of the quiescent class, then the additional fractional field-to-field variance from this random contamination is approximately $\sigma_{\rm rand}^2 \approx p(1-p)/\mu$. Previous work suggests contamination rates of order $\sim10\%$ are plausible for JWST-era photometric quiescent selections, while values as high as $\sim30\%$ have been discussed for less clean selections \citep{AntwiDanso2023,Stevenson2026}. Assuming this random component is independent of the intrinsic cosmic variance, subtracting it in quadrature implies $\sigma_{{\rm CV,int}} \approx (\sigma_{\rm CV}^2-\sigma_{\rm rand}^2)^{1/2}$, or $\approx 0.69$ for $p=0.1$ and $\approx 0.67$ even for the conservative $p=0.3$ case. Thus, purely random, field-independent classification noise is highly unlikely to dominate the observed excess scatter.

More generally, contamination need not act only as a random noise term. Because massive red/quiescent galaxies are generally among the most strongly clustered, highly biased galaxy populations \citep[e.g.,][]{Zehavi2011}, the inclusion of true interlopers (for example due to redshift mis-assignments) would be expected to dilute rather than enhance the clustering signal. In that case, the inferred cosmic variance would be lowered by roughly a factor of $(1+f_{\rm interlopers})^2$ \citep[see, e.g.][]{Williams2011}. Contamination of this kind therefore cannot explain an observed $\sigma_{\rm CV}$ that is larger than the {\sc UniverseMachine} prediction, and may instead imply that our measurement is conservative. In addition, our field-to-field variance is estimated from a finite number (though the largest-to-date) of independent pointings and is therefore still sensitive to small-number statistics and to the presence or absence of rare overdensities. On the mock side, our sSFR-based abundance matching is intentionally rank-based and designed to match only the mean number density; it does not guarantee that the abundance-matched {\sc UniverseMachine} selection reproduces the same halo-occupation mix, satellite fraction, or assembly-dependent correlations as the real quiescent population at fixed abundance.

With these caveats in mind, we discuss the implications of our comparison. An observed cosmic variance that exceeds model predictions may indicate that the observed galaxies trace more strongly biased environments than assumed in the model. For example, this population may preferentially occupy rarer peaks or more massive halos, or exhibit a stronger environmental dependence that enhances small-scale clustering. 

To help narrow down the physical origin of the discrepancy, we perform an additional test: instead of abundance matching in sSFR, we match the observed mean number density using halo mass alone, independent of sSFR, by selecting the $N$ most massive halos in each redshift bin. This selection should, if anything, maximize the expected field-to-field fluctuations in the mocks by isolating the most strongly biased tracers (i.e., the most massive halos),  thus serving as the maximum variance predicted by the {\sc UniverseMachine}. This halo-mass-based abundance-matched sample has slightly higher halo masses\footnote{Here peak halo mass is used, which is defined as the maximum mass the (sub)halo ever reached over its history.}, $10^{12.5\pm0.2}\,M_\odot$, compared to $10^{12.1\pm0.4}\,M_\odot$ for the sSFR-abundance-matched sample. We find that the intrinsic cosmic variance of the halo-mass-matched {\sc UniverseMachine} sample is $\sigma_{\rm CV}\sim0.51$, slightly higher than for the sSFR-abundance-matched sample but still below our current observational constraint. This result suggests that the discrepancy is unlikely to be driven solely by the details of the sSFR-based selection and may instead point to additional environment-dependent effects in the real Universe. For example, the quenching probability may be enhanced via processes analogous to galaxy conformity \citep{Weinmann2006}. Because such effects are not calibrated in the {\sc UniverseMachine} empirical model above $z>1$, if quenching depends partly, or even primarily, on large-scale environment rather than halo mass alone, this could result in a higher effective cosmic variance than predicted by {\sc UniverseMachine}.

\section{Discussion}\label{sec:disc}

\subsection{Implications of the joint measurement of high abundance and high bias}\label{diss:implication}

Our key result is that massive quiescent galaxies at $z>3$ are both unexpectedly abundant and strongly clustered, placing a stringent constraint on models of early quenching. These two observables probe different, but not necessarily independent, aspects of the quenching problem. In a downsizing framework, if the earliest-quenched and/or most massive systems preferentially arise in overdense regions, then the same processes that elevate the quenched number density may also enhance the effective bias of the population. In this sense, the abundance evolution constrains how many galaxies quench, while the clustering constrains where they reside. The joint requirement is therefore demanding: matching the abundance alone is insufficient if quenched systems populate the wrong halos or environments, and matching the clustering alone is not meaningful if the quenched population is too rare.

Our high inferred number densities imply that quenching must operate efficiently within the first $\sim 1$--$2$ Gyr of cosmic time. More broadly, the abundance result indicates that the dominant mechanism must be able to rapidly suppress star formation and maintain low sSFRs despite the intense gas supply expected at these epochs \citep{Tacconi2020}. This poses a significant challenge for current models, since it requires not only early and efficient stellar mass assembly, but also equally rapid truncation of star formation. This early quenching is also directly relevant to the massive quiescent population observed at later times. Stellar-population studies of quiescent galaxies at $z\lesssim2$ have long inferred that at least some of these systems formed the bulk of their stars at $z\gtrsim3$, but their subsequent evolution makes it increasingly difficult to reconstruct the conditions under which they originally quenched. Our number-density measurements instead constrain the abundance of massive galaxies that were already quiescent at these early epochs, and can therefore be compared with the early formation histories inferred for the oldest quiescent galaxies at lower redshift. In this way, the $z>3$ population provides a direct link between the fossil record seen at later times and the epoch when the first massive galaxies actually became quiescent.

Complementing the abundance measurement, the clustering signal constrains how quenching is distributed across the large-scale density field. An elevated clustering would suggest that massive quiescent galaxies preferentially occupy more biased regions, linking the emergence of this population to the broader structure in which their host halos grow. 

We measured the cosmic variance of massive quiescent galaxies at $z>3$, finding a possible excess relative to abundance-matched mock galaxies from UniverseMachine. At present, however, this tension is only at low significance ($\sim 1\sigma$) and, as noted in Section~\ref{diss:future}, will require more than 100 additional independent sightlines to establish robustly. If confirmed, this signal would point to a dependence on large-scale environment in early quenching. One useful framework for interpreting such a trend is galactic conformity, in which quenched systems are associated with an enhanced likelihood of  quenched neighbors \citep[e.g.,][]{Weinmann2006,Kauffmann2013,Hartley2015,Kawinwanichakij2016,Berti2017}. Although the physical origin of conformity at low redshift may not directly carry over to the earlier Universe, a conformity-like signal at $z>3$ would suggest that quenching is shaped by more than halo mass alone. Similarly, if galaxies systematically affect each other in a way that changes the local stellar-to-halo mass relation (in either direction), the clustering signal could be enhanced \citep{Wu2024}. The high bias could also be caused by unmodelled halo assembly effects \citep[assembly bias, see e.g.,][]{Jespersen2022, Chuang2024}.

Taken together, the combination of high abundance and elevated clustering is highly restrictive. It favors scenarios in which the dominant high-redshift quenching pathways are both (i) rapid and effective enough to overcome strong gas replenishment and (ii) coupled to large-scale environment strongly enough to enhance the probability of quenching in the vicinity of already-quenched galaxies. This would be difficult to reconcile with quenching driven solely by internal heating within individual halos, since such a process would not naturally influence neighboring halos. Instead, it points to mechanisms that are both efficient and farther reaching.

The model comparison provides a suggestive clue to the nature of such mechanisms. Among the models considered here, the SAMs generally predict higher quenched abundances than the hydrodynamical simulations at $z\gtrsim 3$, and SHARK and GAEA lie closest to our measurements (Figure~\ref{fig:ab_mod}). A notable shared feature of these two models is the inclusion of an additional radiatively efficient AGN wind/outflow channel beyond low-accretion jet/radio heating. This suggests that a radiatively efficient, quasar-wind-like AGN feedback mode may be an important ingredient for producing the observed abundance of $z>3$ quiescent galaxies in the stellar-mass range ($M_\ast \sim 10^{10}$--$10^{11}\,M_\odot$) probed by our sample. Its effectiveness likely depends on the central gas reservoir: when gas is strongly centrally concentrated, ejective feedback can couple more efficiently to the ISM and more readily expel and/or strongly heat it \citep{Belli2024,Davies2024}. Because massive galaxies at earlier times are more compact \citep{JiGiavalisco2022,vanDerWel2014}, their gas reservoirs may also be more centrally concentrated, making this channel particularly effective at high redshift.

The evidence for excess clustering in our $z>3$ quiescent galaxy sample is also qualitatively consistent with a quasar-wind-driven picture, in which rapid quenching is linked to biased regions of large-scale structure. Ejective feedback can act on short timescales by expelling and/or strongly disrupting the cold-gas reservoir, while quasar triggering is expected to be enhanced in the densest peaks where early massive galaxies assemble. Intriguingly, \citet{Zhu2025} recently reported observational evidence consistent with quasar radiative feedback influencing galaxy growth on intergalactic scales at $z=6.3$, finding suppressed nebular emission in galaxies surrounding the extremely luminous quasar J0100+2802 out to $\sim 1$--$2$ Mpc, comparable to the size of a single NIRCam pointing. While the direct connection between suppressed nebular emission and the long-term quenching of massive systems remains to be established, this result provides a physically motivated example of how quasar episodes at the relevant epochs could plausibly contribute to environmentally correlated suppression in their surroundings. If this interpretation is correct, the quiescent galaxies identified here may also be spatially correlated with UV-luminous galaxies, as expected for the biased environments of rapidly growing high-redshift quasars \citep[e.g.,][]{Costa2014,Costa2019}. Measuring this cross-correlation requires a well-characterized selection function for the UV galaxy population across our heterogeneous fields, and we leave such an analysis to future work.

We emphasize, however, that we are not arguing that quasar-wind-driven feedback is uniquely required by our observations, nor that it is necessarily the dominant quenching mechanism in all massive quiescent galaxies at $z>3$. Rather, we identify it as one plausible scenario that is qualitatively consistent with both the high abundance and the tentative excess clustering measured in our sample.

Other known quenching channels may also contribute, although they do not all naturally explain the possible environmental dependence suggested by our measurements. For example, supernova feedback can regulate star formation, but is generally expected to become less effective in massive systems \citep{Dekel2006}. Major mergers can trigger strong starbursts, but current simulations do not generally find a direct connection between merger events and quenching \citep[e.g.,][]{Rodriguez2019,KurinchiVendhan2024}. Halo-scale quenching provides a more direct connection to environment, since the suppression of gas accretion is tied to halo mass and can produce an environmental dependence of quiescence \citep[e.g.,][]{Cattaneo2006}; however, cold streams are expected to continue penetrating massive hot halos efficiently at high redshift \citep{Dekel2006,Dekel2009}. Importantly, these quenching channels are already included, with different implementations, in many of the models considered here (Section \ref{sec:ab_comp_mod}), yet most still underpredict the abundance of massive quiescent galaxies at $z>3$ (Figure \ref{fig:ab_mod}). Our results may therefore require either modifications to how these known mechanisms operate at early times, an additional quenching channel such as quasar-driven feedback, or both. Regardless, any successful model should reproduce not only the high abundance of early quiescent galaxies, but also their possible preference for overdense environments.

Finally, we note that all references above to ``environment'' or ``environmental dependence'' are intended in the descriptive sense defined in Section~\ref{sec:intro}, namely the large-scale spatial context traced by cosmic variance, rather than as evidence for any specific physical quenching mechanism.

\subsection{The number density of $z\sim7$ quiescent galaxies}\label{diss:zgt7}

JWST NIRSpec spectroscopy now enables absorption line anchored SFH inference for quiescent galaxies at $z\sim4$, yielding strong constraints on formation epochs and quenching timescales. Several confirmed systems show short, intense formation followed by rapid early quenching \citep[e.g.,][]{Nanayakkara2025,deGraaff2025}, including GS-9209 at $z=4.658$ quenched by $z_{\rm quench}\approx6.5$ \citep{Carnall2023} and PRIMER-EXCELS-117560 and 109760 at $z\simeq4.6$ with $z_{\rm quench}\approx6.5-7$ \citep{Carnall2024excel}. If these inferences of $z_{\rm quench}$ are correct, some progenitors of the $z\sim3$ to 5 quiescent population should already be quenched massive systems by $z\sim7$. JWST has confirmed at least one such object at $z\simeq7.3$ \citep{Weibel2025}. Wide area NIRCam surveys provide a direct test of whether this early quenched tail is already present at $z\sim7$.

In our $\sim1000\,{\rm arcmin}^2$ NIRCam dataset, the baseline selection (Section \ref{sec:final_sample}) yields only one candidate with $M_*>10^{10}\,M_\odot$ and $z_{\rm phot}>6.5$ (Figure \ref{fig:median_sed}), suggesting a lower abundance than naive expectations from SFH archaeology. However, our baseline sample intentionally excludes very red sources with $\rm F277W - F444W>1$ to mitigate contamination from LRDs. As mentioned in Section \ref{sec:red_removal}, these removed red objects cluster at $z_{\rm phot}\sim7$ under our \textsc{Prospector} based inference (assuming stellar dominated emission), raising the concern that the cut could also remove genuine $z\gtrsim7$ quiescent galaxies. Indeed, \citet{Hviding2025} find that 60\% of spectroscopically confirmed LRDs satisfy $\rm F277W - F444W>1$, so neither keeping nor removing them is safe with photometry alone: inclusion risks non stellar contamination, while exclusion may suppress the population of interest.

We therefore re-analyze the removed red ($\rm F277W - F444W>1$) objects which are plausibly quiescent by our SFH criteria, requiring (i) $z_{\rm phot}>6.5$ and $M_*>10^{10}\,M_\odot$ in all three \textsc{Prospector} fits and (ii) minimum $\log({\rm sSFR}/{\rm yr}^{-1})<-9.5$ across the fits. This yields 12 galaxies. We refit these galaxies with the \texttt{agn\_blue\_sfhz\_13} templates\footnote{\url{https://github.com/gbrammer/eazy-photoz/tree/master/templates/sfhz}} which include an LRD template from \citealt{Killi2024}. Our goal of this analysis is not classification, but to test whether an LRD like template can explain the photometry comparably well. 

\begin{figure*}
    \centering
    \includegraphics[width=0.727\textwidth]{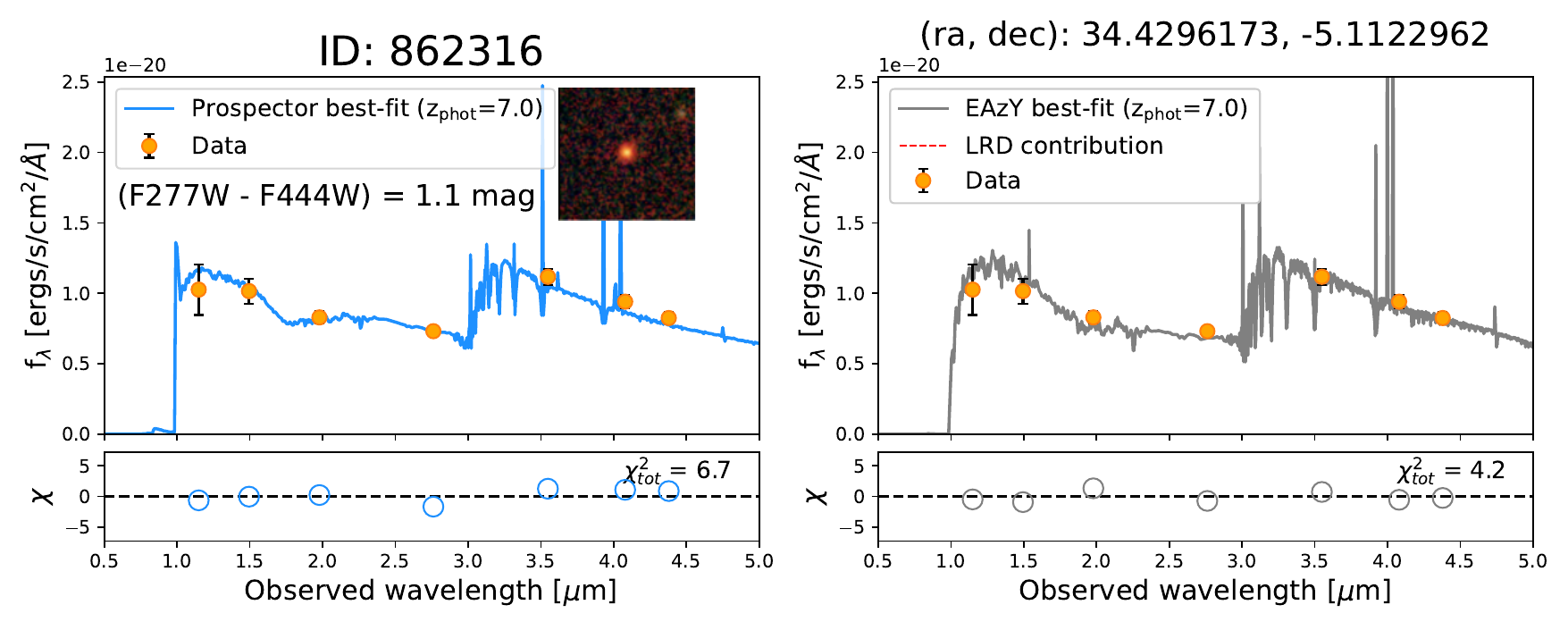}
    \includegraphics[width=0.727\textwidth]{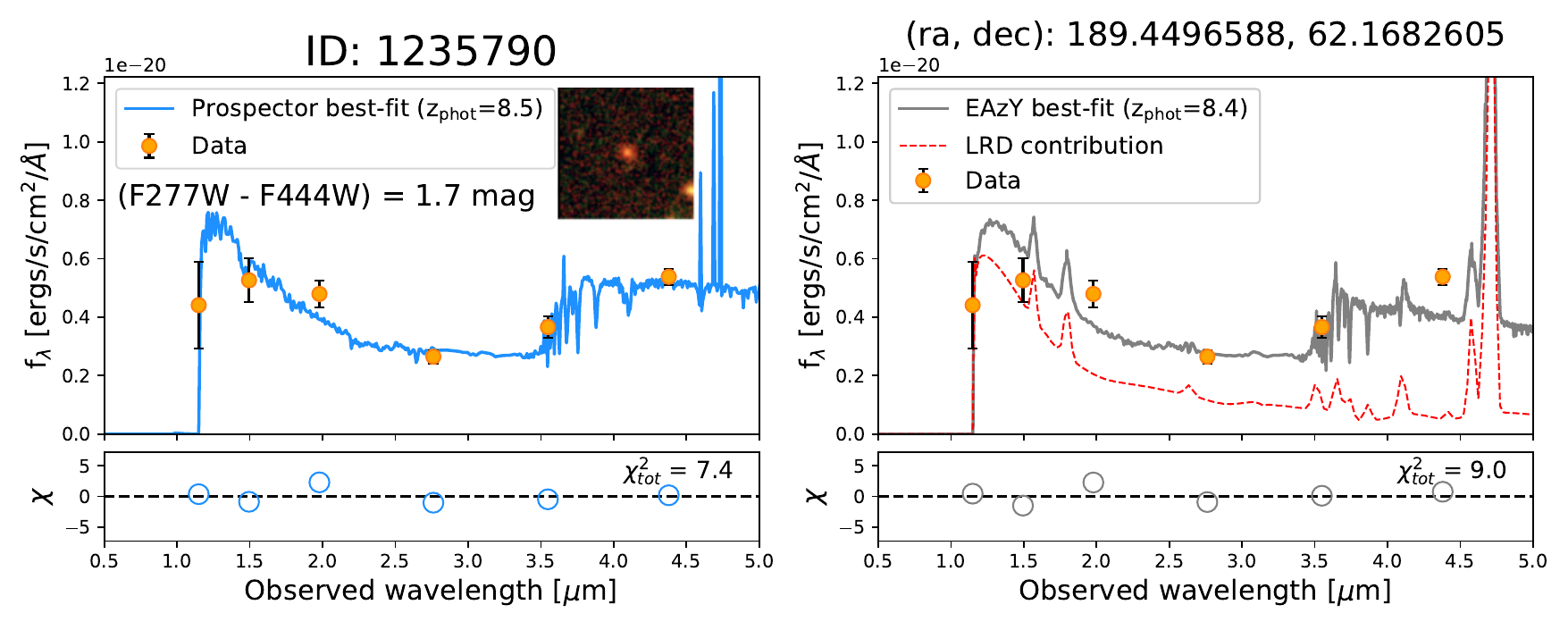}
    \includegraphics[width=0.727\textwidth]{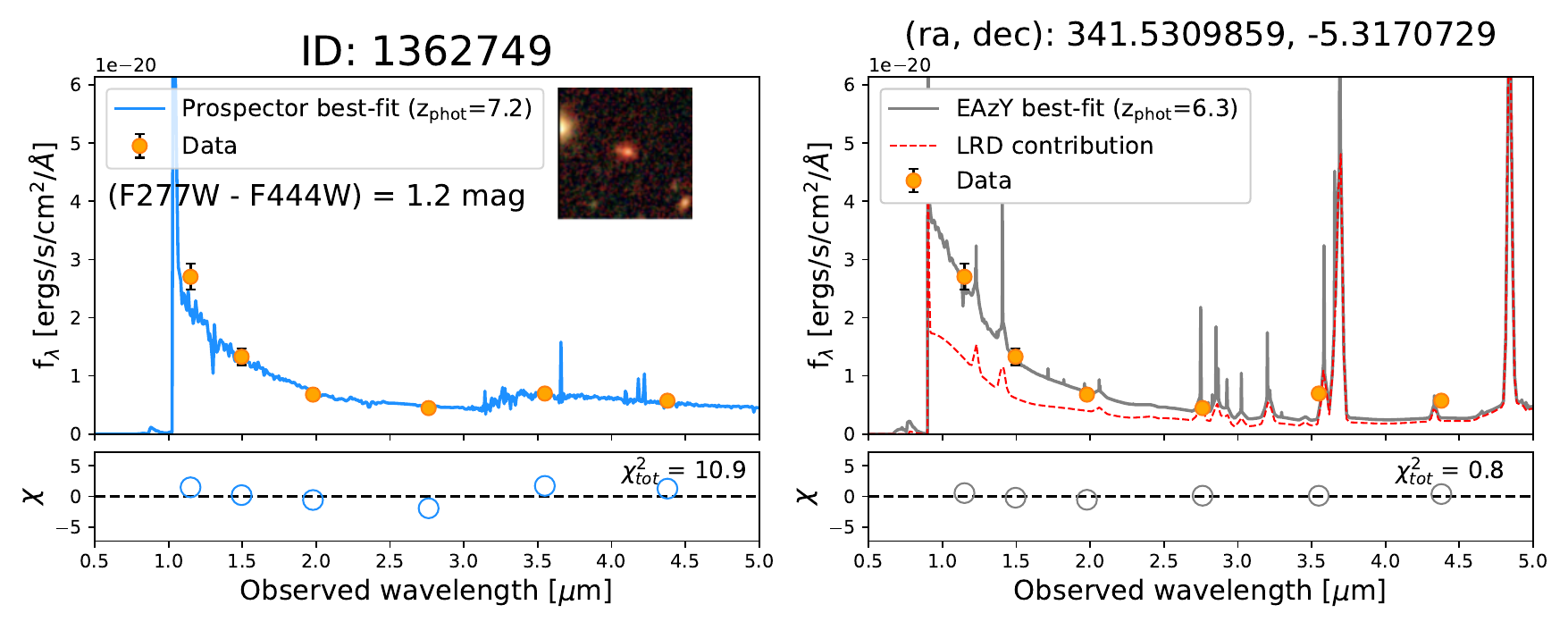}
    \caption{Examples from our re-analysis of the full sample of $z\sim7$ quiescent candidates with $(\mathrm{F277W}-\mathrm{F444W})>1$~mag. For each source, the left panel shows the best-fitting \textsc{Prospector} SED model (no LRD component included) assuming a continuity SFH, with the inset showing the 2"$\times$2" NIRCam cutout, while the right panel shows the best-fitting \textsc{EAzY} template and the dashed red curve indicates the contribution from the LRD component (see Section \ref{diss:zgt7}). The lower panels display normalized residuals and the total $\chi^2$ for each fit. The three rows are representative examples illustrating different outcomes of the re-analysis: from top to bottom, a case where an LRD component is not required, a case where fits with and without an LRD component provide comparably good descriptions of the data, and a case where including the LRD component yields a clearly improved fit.}
    \label{fig:eazy_z7}
\end{figure*}

Figures~\ref{fig:eazy_z7} present examples of this comparison. One source (the top row of Figures~\ref{fig:eazy_z7}) shows negligible LRD contribution and is well fit by both \textsc{Prospector} and EAzY without requiring an LRD component; this is the spectroscopically confirmed $z\simeq7.3$ quiescent galaxy from \citet{Weibel2025}, which was excluded from our initial selection because it has $(\mathrm{F277W}-\mathrm{F444W})=1.1>1$. For 7 sources, the \textsc{Prospector} fit yields a comparable or smaller $\chi^2_{\rm tot}$ than the EAzY stellar+LRD solution, consistent with largely stellar SEDs. For the remaining 13 sources, the EAzY stellar+LRD solution improves $\chi^2_{\rm tot}$ substantially, suggesting plausible LRD-like contributions. We emphasize, however, that these $\chi^2_{\rm tot}$ differences are not decisive. With only 6 to 7 broad bands, distinct physical models can reproduce the same colors, and the EAzY test is incomplete because \texttt{agn\_blue\_sfhz\_13} includes only a single LRD-like template and cannot capture the diversity of LRD SEDs and line properties. Nonetheless, we use this comparison as a pragmatic filter to define an LRD-disfavored subset. 

Retaining only sources for which the EAzY stellar+LRD fit is not substantially better than \textsc{Prospector}, we obtain $N=8$ plausible $z\sim7$ massive quiescent candidates. Over $A=0.28\,{\rm deg}^2$ and $z=6.5$--$8.5$, this implies $n\simeq(1.9\pm1.4)\times10^{-6}\,{\rm Mpc}^{-3}$ (Poisson error only). Systematics from SED modeling and residual LRD or interloper contamination likely dominate, so we treat this as indicative and do not include it in our primary number density results (Section \ref{sec:n}) pending spectroscopic confirmation.

Finally, we compare this provisional $z\sim7$ abundance with expectations from SFH archaeology at $z\sim3$--5. The existence of GS-9209 and the two PRIMER-EXCELS systems, with inferred $z_{\rm quench}\sim6.5$--7, implies that at least three massive quenched galaxies were already in place by $z\sim7$. Although these objects come from targeted spectroscopy and therefore do not have a well-defined survey selection function, adopting a parent area of $A\simeq400\,{\rm arcmin}^2$ for GOODS and UDS yields a crude lower limit of $n\gtrsim2\times10^{-6}\,{\rm Mpc}^{-3}$, comparable to our LRD-disfavored estimate. Current wide-area photometric samples are therefore beginning to test whether the early quenched tail inferred from $z\sim3$--5 spectroscopy is consistent with the observed $z\sim7$ population. Robust progress, however, will require well-defined selection functions and larger, uniformly selected spectroscopic samples; we defer such a comparison to \citet{Zhang2026}.

\subsection{Future strategies for constraining the demographics of $z>3$ massive quiescent galaxies}
\label{diss:future}

Our study demonstrates the power of combining large survey area over many independent sightlines to constrain the evolution of massive quiescent galaxies at $z>3$ using NIRCam. With large area and independent sightlines, we gain not only improved constraints on abundance evolution, but also direct empirical information on clustering through cosmic variance. Together, these observables provide stronger insight into how early massive quiescent galaxies formed. In particular, our results highlight that environment may already play a role in the buildup of quiescent galaxies early in the Universe, if the tentative excess clustering inferred here is confirmed with future data.

\begin{figure}
    \centering
    \includegraphics[width=0.97\linewidth]{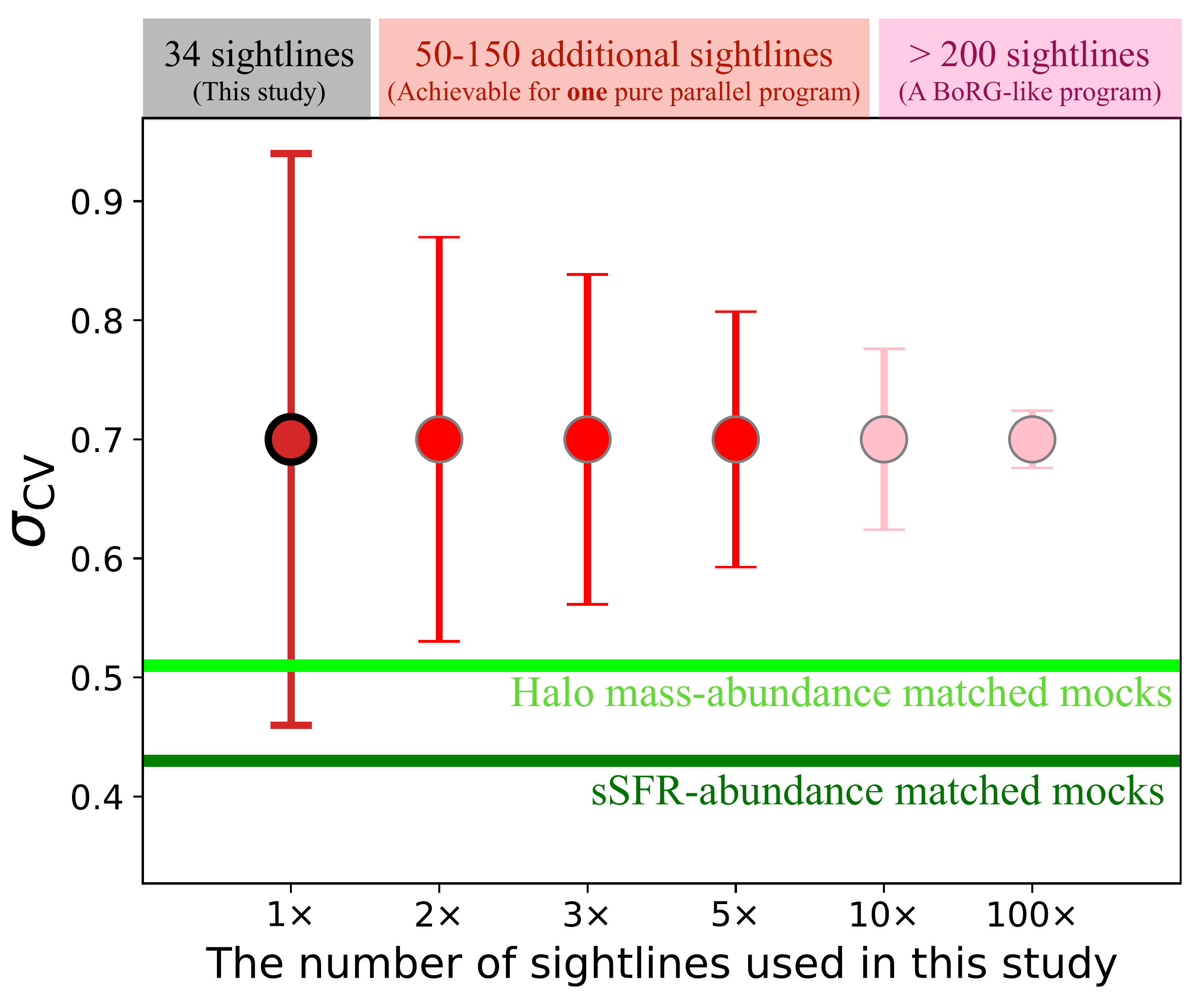}
    \caption{Expected constraints on the cosmic variance, $\sigma_{\rm CV}$, as a function of the number of independent sightlines, in units of the current sample size. The points and error bars show the inferred $\sigma_{\rm CV}$ and its uncertainty assuming the same underlying value as measured in this work, with ``$1\times$'' corresponding to the present dataset of 34 sightlines. The labels at the top indicate approximate observational regimes, from the current study to one pure-parallel JWST program and a BoRG-like multi-cycle survey. The horizontal green lines show the intrinsic cosmic variance of the {\sc UniverseMachine} sSFR-abundance-matched and halo-mass-abundance-matched mock samples (Sections \ref{sec:cv_method_UM} and \ref{sec:cv_results}). Larger numbers of sightlines would substantially improve the ability to distinguish the observed signal from model predictions.}
    \label{fig:cv_sightline}
\end{figure}

For the cosmic-variance measurement, the key quantity is the number of independent sightlines, since the uncertainty on $\sigma_{\rm CV}$ decreases approximately as $1/\sqrt{n_{\rm field}}$ (see Appendix \ref{app:cv_snr}). This makes JWST pure-parallel surveys especially powerful: they naturally produce many single-pointing fields distributed across the sky, enabling direct measurements of field-to-field fluctuations. With 34 independent sightlines, we obtain a first estimate of the cosmic variance of the gold+silver sample and find only mild tension with abundance-matched {\sc UniverseMachine} mocks. Figure~\ref{fig:cv_sightline} shows how this test improves with increasing sightline number: while the present dataset still yields broad uncertainties, adding $\sim$50--150 independent pointings, achievable within a single pure-parallel program with $\gtrsim6$ NIRCam filters \citep{jwpure}, would already tighten the constraint substantially. Importantly, the massive quiescent galaxies targeted here are several AB magnitudes brighter than typical pure-parallel depths.

Looking further ahead, the accumulation of hundreds of sightlines over multiple JWST cycles \citep[e.g., analogous to superBoRG, which assembled $\gtrsim300$ independent HST parallel pointings collected over a decade;][]{Morishita2021} would not only tighten variance-based tests dramatically, but could also begin to build statistical samples of candidate quiescent galaxies at $z\gtrsim7$. This is difficult to achieve with wide-area near-infrared surveys such as Roman or Euclid, because the selection of these systems requires well-sampled photometry beyond $3\,\mu{\rm m}$, which at present is only available with JWST.

We also emphasize that filter completeness is at least as important as depth for quiescent-galaxy classification. For red, low-sSFR systems at $z>3$, the main failure mode is often not missing faint sources, but misclassifying them because the SED is poorly constrained around the Balmer/4000\,\AA\ break. In practice, continuous wavelength coverage across the short- and long-wavelength NIRCam channels is critical: removing F200W, for example, creates a gap between F150W and F277W that worsens photometric-redshift stability and increases the model dependence of SED fitting. Future surveys optimized for quiescent-galaxy demographics should therefore prioritize filter sets that minimize such gaps, even at some cost in single-band depth.

Finally, spectroscopy remains the bottleneck for turning candidates into robust constraints. Many JWST/NIRSpec confirmations come from heterogeneous programs that target relatively small, pre-selected samples \citep[e.g.,][]{Nanayakkara2024,Carnall2024,Baker2025b}. Their complex and non-uniform selection functions complicate attempts to derive an unbiased census. Recent JWST spectroscopy has also begun to address this limitation. In particular, RUBIES \citep{deGraaff2025} has explicitly attempted spectroscopic completeness to enable abundance based on spectroscopically confirmed samples \citep{Zhang2025}. Nevertheless, cosmic variance remains a key limitation for any program that target 
a small number of contiguous fields.
A natural next step is therefore large spectroscopic programs that target uniformly selected samples across many independent sightlines, which can (i) quantify contamination rates in ``silver''-like selections, and (ii) directly measure quiescence diagnostics (e.g., Balmer absorption and continuum breaks) while robustly constraining the star-formation histories of these galaxies.

\section{Summary}

In this work, we present a wide-area  JWST/NIRCam study designed to measure both the cosmic abundance and field-to-field variance of massive quiescent galaxies at $z \gtrsim 3$. Combining the JWST/NIRCam pure-parallel survey PANORAMIC and several legacy surveys in deep fields, we assemble the \emph{largest JWST/NIRCam imaging dataset to date for this purpose}. Restricting to regions with $\geq 6$ NIRCam filters, we compile a $\sim 0.28~\mathrm{deg}^2$ ($\sim 1000~\mathrm{arcmin}^2$) dataset distributed across $\sim 34$ independent sightlines. This scale enables number-density measurements that are less skewed by rare environments and less biased by small footprints, providing a direct empirical constraint on cosmic variance.

Using a carefully identified sample of quiescent galaxy candidates spanning $z \approx 3$--8 with $M_\star \ge 10^{10}~M_\odot$, we infer number densities using a probabilistic approach. We find that the abundance of massive quiescent galaxies declines steeply with redshift: the conservative gold sample drops from $\sim 1.8\times10^{-5}~\mathrm{Mpc}^{-3}$ at $z=3.0$--3.5 to $\sim 0.7\times10^{-6}~\mathrm{Mpc}^{-3}$ at $z\sim6$, with the combined gold+silver sample yielding different absolute densities but a consistent evolutionary trend. These results show that massive quenched systems were already in place in the early Universe, but became progressively less common toward $z\gtrsim6$. At $z\sim7$, we identify a small set of plausible candidates, implying a number density of order $10^{-6}~\mathrm{Mpc}^{-3}$, but this estimate remains provisional because photometry alone cannot securely distinguish such systems from LRD/AGN contaminants.

We compare our measured number densities to predictions from a suite of SAMs and hydrodynamical simulations, finding that they underpredict the measured abundance of massive quiescent galaxies at $z\gtrsim 4$ by $\gtrsim 1$ dex, highlighting persistent challenges in reproducing both early quenching. We separately assess {\sc UniverseMachine} as an empirically calibrated baseline, anchored primarily at lower redshift and extrapolated to early times. The growing tension between {\sc UniverseMachine} and the observed abundance of massive quiescent galaxies at $z\gtrsim 5$ suggests that reproducing our measurements may require earlier and/or more efficient quenching (and/or maintenance of quiescence) than implied by relations calibrated at lower redshift.

The multi-sightline NIRCam coverage enables an empirical measurement of cosmic variance. We measure large field-to-field variance, with $\sigma_\mathrm{CV}\approx 0.7\pm0.3$. We compare this observed cosmic variance  to abundance-matched {\sc UniverseMachine} mocks, constructed to match the observed mean abundance. The mocks exhibit systematically smaller field to field variance, 
providing tentative evidence (large errorbars) that the observed quiescent galaxy population is 
more strongly clustered (i.e., more highly biased) than {\sc UniverseMachine} predicts once the mean abundance is matched.

Taken together, our findings favor quenching pathways that are both rapid enough to overcome strong gas replenishment at early times and sufficiently coupled to large-scale environment to enhance the probability of quenching in biased regions. More broadly, these results suggest that quenching driven solely by internal heating within individual halos may be insufficient to explain the observed combination of high abundance and strong clustering at $z>3$.

Finally, this work demonstrates the power of large-area, multiple-sightline surveys for studying rare, highly biased populations in the early universe. For massive quiescent galaxies at $z > 3$, increasing the number of independent sightlines is essential not only for reducing the influence of rare structures on abundance measurements, but also for directly characterizing their large-scale distribution. The dataset presented here marks a first step in that direction: it shifts the $z > 3$ quiescent-galaxy census from field-limited measurements to a variance-controlled census. We are still only at the beginning. As larger imaging and spectroscopic samples extend this approach across many more sightlines, they will turn early quenched galaxies into a powerful laboratory for understanding how the first massive galaxies formed, quenched, and traced structure in the young universe.

\section*{Acknowledgments}

This work is based in part on observations made with the NASA/ESA/CSA James Webb Space Telescope. The data were obtained from the Mikulski Archive for Space Telescopes at the Space Telescope Science Institute, which is operated by the Association of Universities for Research in Astronomy, Inc., under NASA contract NAS 5-03127 for JWST. These observations are associated with program JWST-GO-2514. ZJ, CCW, and KEW acknowledge support for program JWST-GO-2514.  Support for program JWST-GO-2514 was provided by NASA through a grant from the Space Telescope Science Institute, which is operated by the Association of Universities for Research in Astronomy, Inc., under NASA contract NAS 5-03127.  The work of CCW is supported by NOIRLab, which is managed by the Association of Universities for Research in Astronomy (AURA) under a cooperative agreement with the National Science Foundation. This work has received funding from the Swiss State Secretariat for Education, Research and Innovation (SERI) under contract number MB22.00072, as well as from the Swiss National Science Foundation (SNSF) through project grant $200020\_207349$. The Cosmic Dawn Center (DAWN) is funded by the Danish National Research Foundation under grant DNRF140. This work was performed in part at Aspen Center for Physics, which is supported by National Science Foundation grant PHY-2210452. P. Dayal warmly acknowledges support from an NSERC discovery grant (RGPIN-2025-06182).

\bibliography{sample7}{}
\bibliographystyle{aasjournal}

\appendix 

\section{Comparison of photometry} \label{app:compare_photometry}

\begin{figure*}
    \includegraphics[width=1\textwidth]{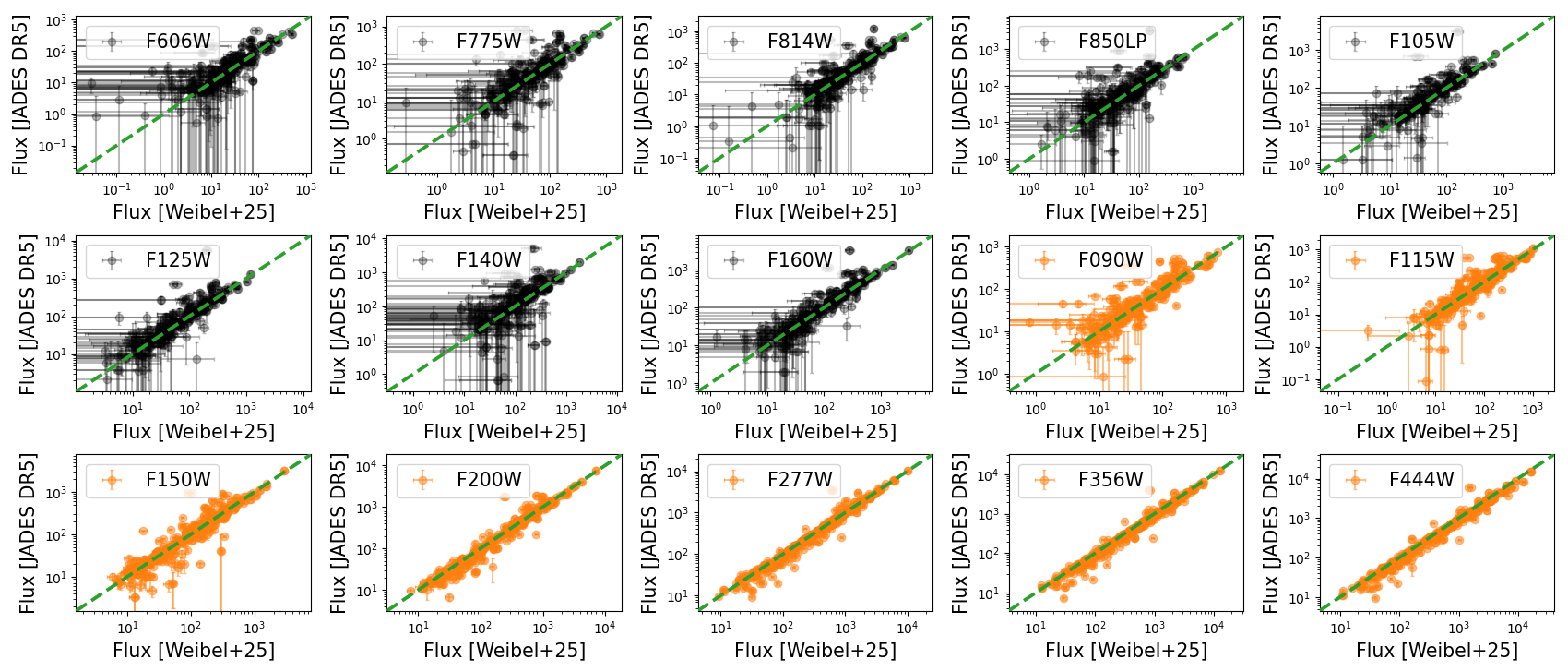}
    \caption{HST (black) and NIRCam (orange) photometric comparison for the initially selected galaxies in GOODS-S. The $x$-axis shows the photometry from our pipeline (Section \ref{sec:data}), and the $y$-axis shows the measurements from the latest JADES DR5. The fluxes are given in units of nJy. The green dashed line marks the one-to-one relation. The two sets of photometry are highly consistent within the uncertainties.}
    \label{fig:compare_phot}
\end{figure*}

To assess the quality of our photometric procedure described in Section \ref{sec:data}, we compare the HST and JWST/NIRCam photometry derived from our pipeline for the initially selected sample galaxies (Section \ref{sec:initial_selection}) in GOODS-S  with the Kron photometry released by the JADES team \citep[DR5;][]{Johnson2026,Robertson2026}. We note that the JADES imaging reduction and photometry were performed with a customized GTO pipeline, and that their source detection and segmentation strategies differ somewhat from those used in the {\sc grizli} pipeline adopted in our study (Section \ref{sec:data}). Nonetheless, as shown in Figure \ref{fig:compare_phot}, our photometry is highly consistent with that from JADES within the uncertainties across both HST and NIRCam, indicating that, for the galaxy sample considered in this work, the results are not sensitive to the choice of pipeline. This consistency demonstrates the robustness of our measurements.

\section{SED fitting assumptions} \label{app:sed_assumption}

We employed the FSPS stellar population synthesis code \citep{Conroy2009, Conroy2010} with MIST stellar isochrone libraries \citep{Choi2016, Dotter2016} and MILES stellar spectral libraries \citep{Falcon-Barroso2011}. We applied the \citet{Madau1995} IGM absorption model and included the nebular emission model of \citet{Byler2017}. Dust attenuation was modeled following \citet{Tacchella2022}, in which the attenuation of nebular emission and young stellar populations was treated differently from that of old stellar populations \citep{Charlot2000}. The redshift was fit as a free parameter with a flat prior of $z\in(0,20)$. In this work, we carried out SED fitting with three SFH models that are commonly employed in the literature:
\begin{itemize}
    \item Parametric delayed–$\tau$ SFH -- SFR(t) $\propto te^{-t/\tau}$.
    \item Non-parametric SFH with the continuity prior \citep{Leja2019} modeled as a piecewise step function composed of nine lookback time bins. The first two bins are fixed to $0-30$ and $30-100$ Myr;  the last bin is fixed to $0.85t_{\rm{H}}-t_{\rm{H}}$ where $t_{\rm{H}}$ is the Hubble time at the epoch of observation; the remaining six bins are evenly spaced in logarithmic time between 100 Myr and $0.85t_{\rm{H}}$. Changes in SFR between adjacent bins, $x = \log(\rm{SFR_{t_i}/SFR_{t_{i+1}}})$, are modeled as a Student’s $t$-distribution with $\sigma = 0.3, \nu = 2$ (see \citealt{Leja2019} for details), which strongly disfavors rapid changes in SFR over short timescales.
    \item Non-parametric SFH with the bursty continuity prior -- Similar to the continuity prior SFH, but adopting $\sigma = 1$ in the Student's $t$-distribution, thereby permitting more bursty star formation \citep[e.g.,][]{Tacchella2022}.
\end{itemize}

\begin{figure*}
    \includegraphics[width=1\textwidth]{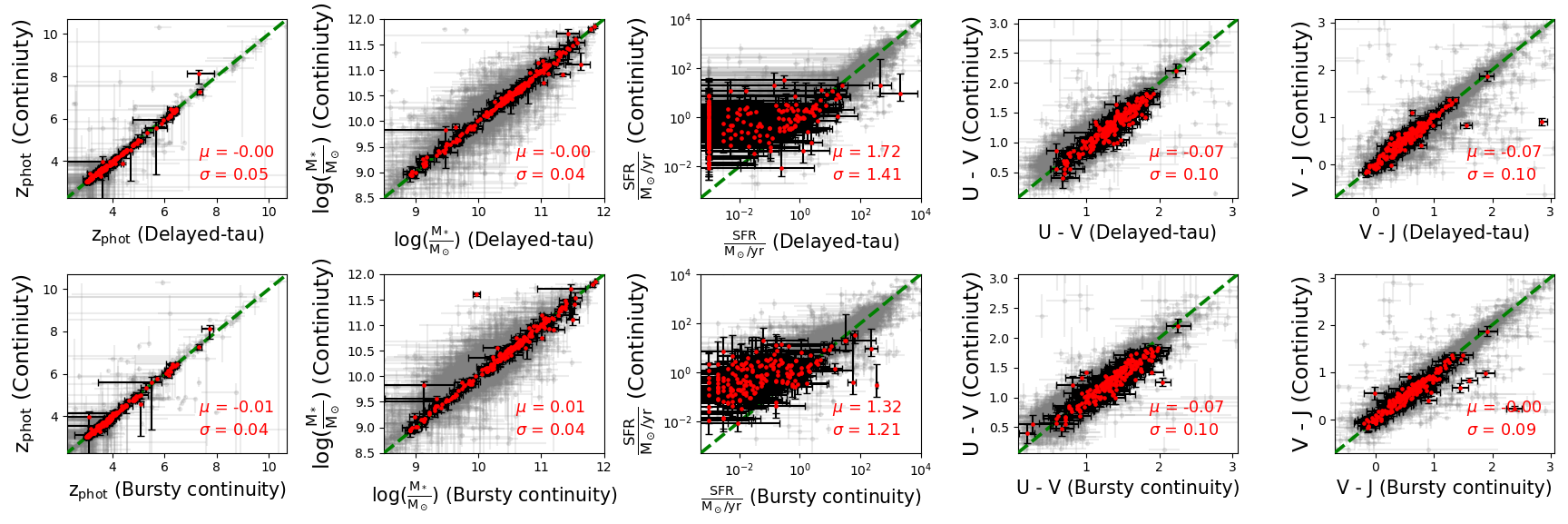}
    \caption{Comparisons of the inferred properties assuming three different SFHs (Section \ref{sec:refine_selection})  for the initial sample of 2633 galaxies (Section \ref{sec:initial_selection}). The green dashed line marks the one-to-one relation. For clarity, galaxies with SFR $<10^{-3}$ M$_\sun$ yr$^{-1}$ are all plotted at $10^{-3}$ M$_\sun$ yr$^{-1}$. Galaxies with sSFR $\le10^{-10}$ yr$^{-1}$ from the non-parametric continuity prior are highlighted in red, showing the large systematic uncertainties in SFR measurements for quiescent galaxies when relying on photometric data alone. In each panel, we also label the median ($\mu$) and standard deviation ($\sigma$) of $y-x$ for these low-sSFR galaxies. }
    \label{fig:sed_compare}
\end{figure*}

In Figure \ref{fig:sed_compare}, we compare $z_{\rm phot}$, M$_*$, SFR, and rest-frame (U$-$V) and (V$-$J) colors derived from the three SED runs for the initially selected sample. Except for SFR, all inferred properties are highly consistent across models. At the high-SFR end, all three SFH models produce consistent results; however, for galaxies with low sSFR, systematic offsets become apparent. The inferred SFR can differ substantially between models, underscoring the significant uncertainties associated with constraining the SFR of quiescent galaxies using photometric data alone.

We compare the goodness of fit among the three SFH models using $\chi^2_{\rm tot}$, defined as the $\chi^2$ summed over all available photometric bands.\footnote{For the best-fit parameters, we adopt the median values of the posterior distribution. We also verify the results using the maximum-likelihood solution and find negligible differences.} We find that approximately 60\% of our quiescent sample favors -- i.e., has the smallest $\chi^2_{\rm tot}$ -- the delayed-$\tau$ SFH, while roughly 25\% and 15\% favor the bursty continuity and continuity SFHs, respectively. These fractions are not sensitive to the number of available filters. This result is consistent with \citet{Helton2025}, who reported that SED fitting with a delayed-$\tau$ SFH tends to better reproduce high-redshift galaxy colors than other SFHs. 

We also note that the three SFH models do not contribute equally to the Silver sample. Among Silver galaxies with $M_*\ge 10^{10}M_\odot$, the bursty-continuity fit satisfies the sSFR criterion (Equation \ref{equ:ssfr_thresh}) for 84\% of the sample, compared with 38\% for the delayed-$\tau$ fit and 21\% for the standard continuity fit. For 55\% of the Silver sample, the bursty-continuity fit is the only one that satisfies the criterion. This trend does not change strongly with redshift, with corresponding bursty-continuity fractions of 87\%, 75\%, and 81\% at $z=3$-4, 4-5, and 5-8, respectively. This difference mainly reflects the adopted SFH priors. The bursty-continuity prior allows much larger changes in SFR between adjacent time bins and can therefore more easily produce a rapid decline in the most recent SFR, while the standard continuity prior disfavors such rapid changes. The delayed-$\tau$ model gives an intermediate case, since a sufficiently short $\tau$ can also produce a low present-day sSFR through the overall decline of the parametric SFH.

One concern is that our SED fitting does not include an AGN component, so inferred galaxy properties could be biased if AGN are present. Although we exclude LRD-like sources (Section \ref{sec:red_removal}), other AGN (e.g., X-ray and mid-IR) may remain. We estimate the impact using GOODS-S, where ultradeep Chandra data \citep{Luo2017} and JWST/MIRI observations \citep{Alberts2024smiles,Rieke2024} are available. Cross-matching with the GOODS-S AGN catalog of \citet{Lyu2024}, which uses multiwavelength (X-ray to radio) criteria, we find that 4 of 42 ($\approx9.5\%$) massive ($>10^{10}\,M_\odot$) galaxies in our gold+silver samples are classified as AGN. Using the AGN-inclusive SED fits from \citet{Lyu2024}, we find that the rest-frame UV–NIR SEDs of these four galaxies are still dominated by stellar emission. Prior work also suggests that for non-quasar AGN, omitting AGN components mainly adds scatter to $M_*$ and introduces $\sim0.2$ dex uncertainty in SFR, with no strong systematic bias \citep{Ji2022agn,Santini2012,Yang2018}. This is unlikely to change our quiescent classification, since our galaxies lie $\gtrsim0.5$ dex below the star-forming main sequence (Figure \ref{fig:sfms}). Given the low inferred AGN fraction ($\lesssim10\%$) and the modest expected biases, residual AGN should have only a minor impact on our results.

\section{Removed sources from visual inspection}
\label{app:visual_removal}

A total of 20 sources were removed from the analysis based on our visual inspection of the imaging quality. The majority (13) of the defected sources are located in the PANORAMIC pointing J153500, where very bright sources in the F200W mosaic exhibited central flux deficits. This effect was likely caused by the  short  exposure times which resulted in poorer-than-usual cosmic-ray rejection \citep{Williams2025}. Those authors updated the pixel-flagging procedure in J153500 to implement more aggressive cosmic-ray rejection, albeit at the cost of image quality for the brightest sources. As a result, some bright sources showed artificially reduced F200W flux in their central pixels, which in turn produced an apparent flux discontinuity between the F200W and F277W filters, mimicking the appearance of a Balmer break at $z \sim 6$ in the photometry. This effect led to the false identification of those 13 quiescent galaxy candidates in this pointing.

We note that our initial candidate selection following \citet{Long2024} did not rely on F200W (Equation \ref{equ:red_wedge}). Thus, the reduced F200W image quality in J153500 did not raise concerns about missing high-redshift candidates at that stage. For our refined selection based on SED fitting, however, a concern remains that legitimate $z>3$ quiescent galaxies in this pointing might have been erroneously excluded owing to problematic F200W photometry. To assess this, we re-ran our SED analysis for the 32 initially selected sources in this pointing, excluding their F200W fluxes from the {\sc Prospector} fitting. No new robust quiescent galaxies were identified. We are therefore confident that all qualified candidates in this pointing have been recovered, and since fainter candidates were not affected by this pixel flagging during processing, we include this pointing and its robust candidates in the subsequent analysis.

From this test with the 5-filter NIRCam set (i.e., excluding F200W), we also found that the quality of the SED fitting decreases significantly. In particular, the large gap between F150W and F277W leads to an increase in the standard deviation of the $z_{\rm{phot}}$ measurement for our sample to 0.7 between the continuity and delayed-$\tau$ SFHs, and to 0.5 between the continuity and bursty-continuity SFHs -- both substantially larger than the value ($\sim$0.05) found with the 6-filter NIRCam set (Figure \ref{fig:sed_compare}), which is the baseline requirement of this work (Section \ref{sec:data}).

\section{Cross-matching with spectroscopically confirmed quiescent galaxies at $z>3$ from the literature} \label{app:crossmatch}

We cross-matched our final samples (Section \ref{sec:final_sample}) with spectroscopically confirmed quiescent galaxies at $z \gtrsim 3$ from two recent studies: \citet{AntwiDanso2025} and \citet{Zhang2025}. \citet[][their Table 3]{AntwiDanso2025} compiled a list of quiescent galaxies at $z>3$ spectroscopically confirmed through both ground-based and space-based observations (see references therein). In addition, \citet[][their Table 2]{Zhang2025} reported quiescent galaxies from the RUBIES program confirmed with JWST/NIRSpec spectroscopy.
Among the 79 galaxies reported, we found that 35 (after removing duplicates) lie at $z \geq 3$ and fall within our imaging coverage with 6-band NIRCam photometry. 
Since the list of \citet{AntwiDanso2025} includes a few galaxies from earlier studies based on low-S/N ground-based spectroscopy, we excluded ZF-COS-20133, ZF-COS-17779, and ZF-COS-19589, all of which were flagged as having uncertain spectroscopic confirmations in the original paper \citep{Schreiber2018}. After these removals, we were left with a total of 32 spectroscopically confirmed quiescent galaxies at $z > 3$ for cross-matching.  29 are recovered by our selection procedure, with 19 in the gold sample and 10 in the silver sample. Four galaxies  were not ultimately selected and we examined the reasons for their exclusion:
\begin{itemize}
    \item 3D-EGS-26047 at $z_{\rm spec}=3.234$ and 39138 at $z_{\rm spec}=3.442$, both included in the sample of \citet{AntwiDanso2025}, were initially selected into our parent sample. In our three SED fits, both sources lie about 0.3--0.5 dex below the star-forming main sequence at $z\sim3.5$. However, neither source is retained in our final sample, because both fail to satisfy the sSFR criterion adopted in Section~\ref{sec:refine_selection}, i.e. $\mathrm{sSFR}\lesssim10^{-10}\,\mathrm{yr}^{-1}$ at these redshifts.

    We also examine the available spectroscopic information for these two sources. For 3D-EGS-26047, the Keck/MOSFIRE spectrum is relatively noisy and does not cover H$\alpha$, but tentative detections of [\ion{O}{3}]$\lambda5007$, H$\beta$, and [\ion{O}{2}]$\lambda3727$ suggest a younger stellar population than is typical of quiescent systems. For 39138, JWST/NIRSpec prism spectroscopy shows a clear emission feature at the wavelength of H$\alpha$, although the low spectral resolution prevents the [\ion{N}{2}] contribution from being separated. \citet{Jin2024} therefore conservatively estimated $\mathrm{sSFR}=10^{-9.9}\,\mathrm{yr}^{-1}$, placing this source very close to the selection boundary adopted here.

     \item UNCOVER-18407 at $z_{\rm{spec}} = 3.97$ \citep{Setton2024} was {\it not} initially selected, as it is marked with a \texttt{hugekron\_flag} in the catalog of \citet{Weibel2025uvlf}. Sources with this flag were excluded from the outset (see the beginning of Section \ref{sec:selection}).  Inspection of its image cutouts reveals four bright, very extended galaxies within $\lesssim2''$ of its position, which likely caused problematic segmentation in the automatic pipelines.
     
\end{itemize}

\begin{figure*}
\centering
    \includegraphics[width=0.97\textwidth]{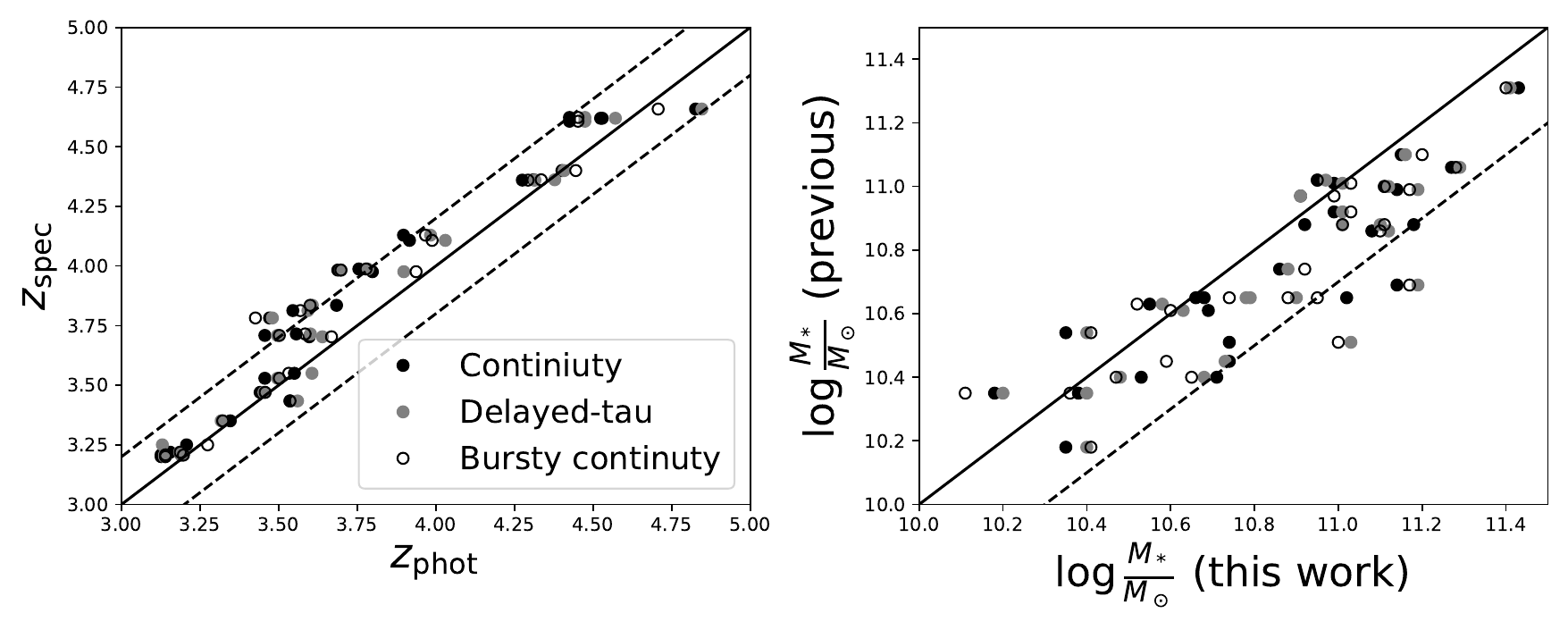}
    \caption{Comparison of our SED-derived properties with those reported in the literature for the 28 spectroscopically confirmed quiescent galaxies at $z>3$. Results assuming the continuity, delayed-$\tau$, and bursty-continuity SFHs are shown as black and grey filled circles, and black open circles, respectively. The black solid line indicates the one-to-one relation. {\bf Left:} photometric versus spectroscopic redshifts, with the two black dashed lines marking $\Delta z = \pm0.2$. {\bf Right:} stellar masses from our SED fittings compared to those reported in the literature. The black dashed line marks $\Delta(\log M_*)=-0.3$ dex. We note that stellar-mass estimates in previous works are based on different methods and data sets, and thus are not directly homogeneous.}
    \label{fig:sed_compare_spec}
\end{figure*}

Finally, for the 29 matched galaxies, we compare our SED-derived redshifts and stellar masses with those reported in the literature, as shown in Figure \ref{fig:sed_compare_spec}. For redshifts, our photometric values are in excellent agreement with spectroscopic measurements, with nearly all sources lying within $|\Delta z|\lesssim0.2$. For stellar masses, we also find strong agreement, although our estimates are systematically higher by $\sim0.1$–0.2 dex on average. Such offsets are well known in the literature and are commonly attributed to systematic differences between SED modeling methods and assumptions \citep{Leja2019,Lower2020,Ji2022,Ji2023,Leja2022, Jespersen2025c}.

\section{Cosmic-variance analysis for the gold sample} \label{app:gold}

\begin{figure*}
    \includegraphics[width=0.347\textwidth]{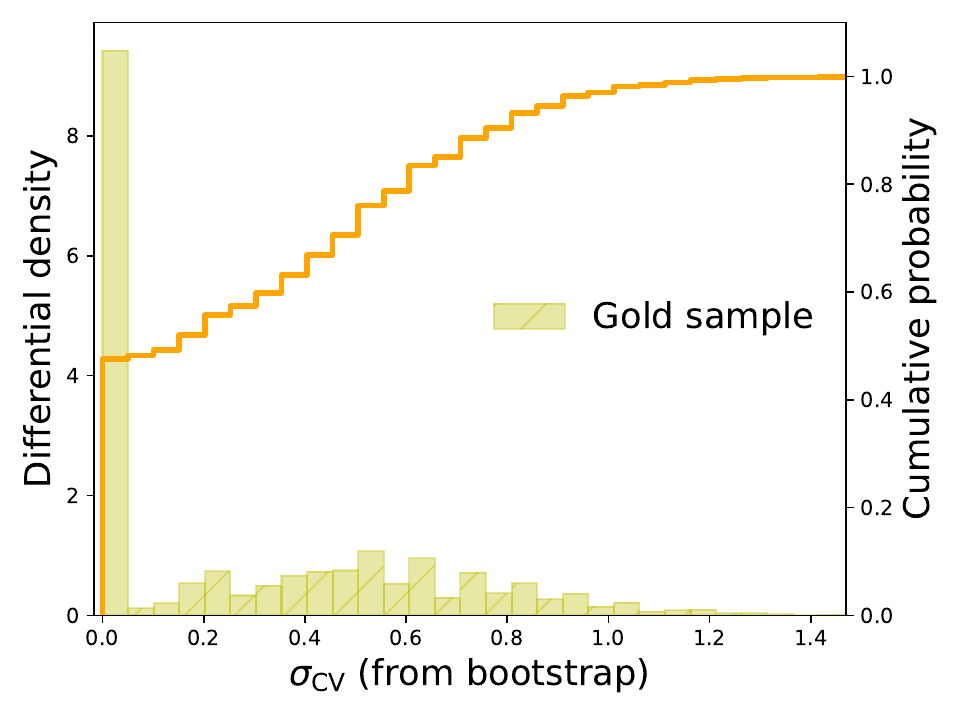}
    \includegraphics[width=0.347\textwidth]{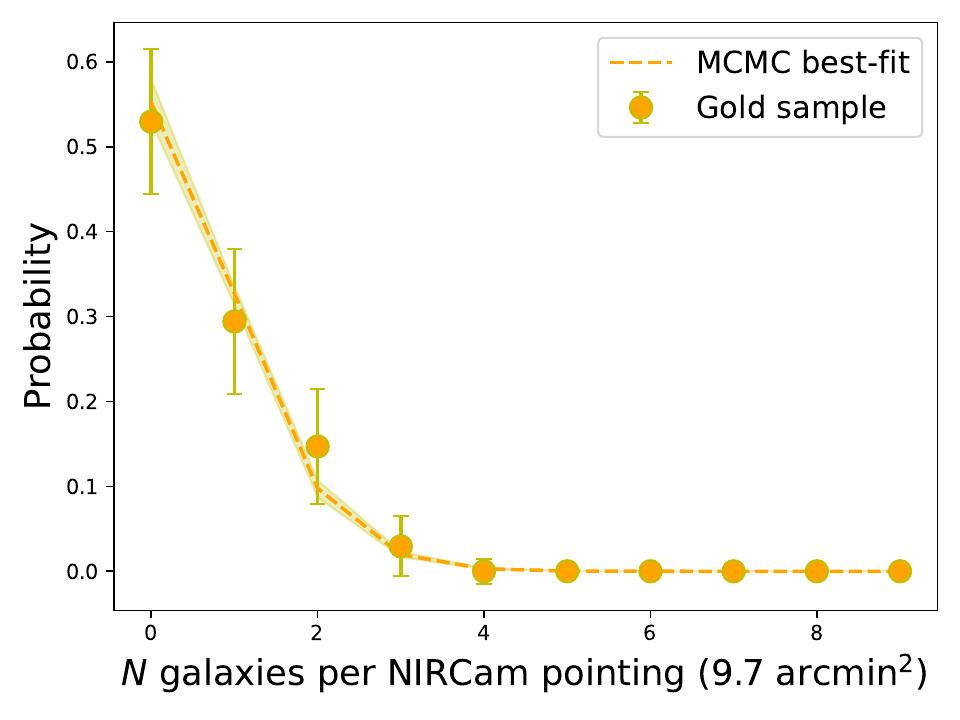}
    \includegraphics[width=0.27\textwidth]{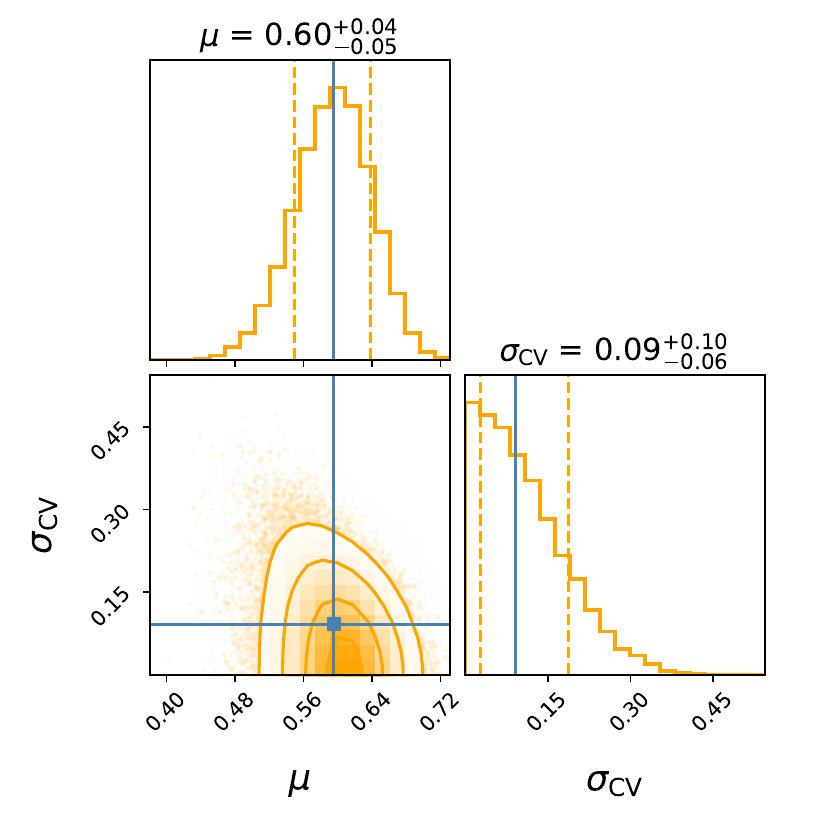}
    \caption{Left panel is similar to Figure~\ref{fig:cv_com_boots}, but shows the bootstrap-inferred $\sigma_{\rm CV}$ for the gold sample. Approximately 50\% of the bootstrap realizations return $\sigma_{\rm CV}=0$. The middle and right panels are similar to Figure~\ref{fig:cv_com}, but show the MCMC-fitting inferred $\sigma_{\rm CV}$ for the gold sample; the $\sigma_{\rm CV}$ posterior ``piles up'' at zero. Together, these results indicate that the gold sample alone lacks sufficient statistics for a robust cosmic-variance measurement.}
    \label{fig:cv_gold}
\end{figure*}

Here we assess whether the gold sample alone provides sufficient statistics to constrain the cosmic variance on NIRCam-pointing scales. We apply the same two estimators used in the main text, namely, bootstrap variance decomposition and Bayesian distribution fitting, but restricting the analysis to the gold sample only. As Figure~\ref{fig:cv_gold} shows,  the inferred $\sigma_{\rm CV}$ using the bootstrap method frequently reaches the physical boundary at zero: $\sim50\%$ of bootstrap realizations return $\sigma_{\rm CV}=0$. This indicates that the small number of gold objects and the limited number of independent pointings prevent a measurable excess variance from being robustly detected. Consistent with this, the MCMC distribution-fitting approach yields a posterior for $\sigma_{\rm CV}$ is strongly concentrated toward $\sigma_{\rm CV}=0$, implying that the gold-sample counts distribution does not require super-Poisson dispersion. 
We therefore conclude that the gold sample by itself does not contain sufficient statistical power for a robust cosmic-variance measurement, and we adopt the combined gold+silver sample for the main cosmic-variance analysis presented in Section~\ref{sec:cv}.

\section{Detectability of Cosmic Variance}
\label{app:cv_snr}

Because the measured cosmic variance depends not only on the intrinsic clustering signal but also on the finite number of independent sightlines, it is useful to show explicitly how its detectability scales with survey design. For rare objects, the field-to-field variance is often dominated by Poisson fluctuations, making it difficult to distinguish a nonzero cosmic-variance term from noise. The purpose of this appendix is therefore not to provide a rigorous statistical estimator, but rather to show the approximate scaling of the detectability of $\sigma_{\rm CV}$ with the mean number of objects per sightline and the number of independent fields.

Let $N$ denote the number of galaxies per independent sightline, and let $s_N^2$ be the measured field-to-field sample variance across the $n_{\rm field}$ sightlines, which estimates the underlying variance ${\rm Var}(N)$. Similar to Equation~\ref{eq:var_mu_sigcv}, we write
\begin{equation}
\sigma_{\rm CV}^2 = \frac{{\rm Var}(N)-\langle N\rangle}{\langle N\rangle^2},
\end{equation}
where \(\langle N\rangle\) is the mean number of objects per sightline. Replacing the underlying variance by its sample estimator \(s_N^2\)\footnote{\(s_N^2\) is the unbiased sample-variance estimator of the field-to-field variance, i.e., \(\mathbb{E}[s_N^2]={\rm Var}(N)\).}, the corresponding estimate becomes
\begin{equation}
\sigma_{\rm CV}^2 \approx \frac{s_N^2-\langle N\rangle}{\langle N\rangle^2}.
\end{equation}
Assuming the field-to-field count distribution is  Gaussian, the variance of the sample variance is
\begin{equation}
{\rm Var}(s_N^2)\approx \frac{2\,{\rm Var}(N)^2}{n_{\rm field}-1}
\end{equation}
\citep[e.g.,][]{casella2002statistical,degroot2012probability}.
Note that this assumption is adopted only to derive the approximate scaling relation; non-Gaussianity would modify the prefactor\footnote{If the field-to-field count distribution is a Gamma function \citep{Jespersen2025}, we can show that Equation~F3 becomes
$\mathrm{Var}(s_N^2)\approx \left[\frac{2}{n_{\rm field}-1}+\frac{6}{n_{\rm field}}\left(\frac{1}{\mu}+\sigma_{\rm CV}^2\right)\right]\mathrm{Var}(N)^2
$. This reduces to the Gaussian result in the limit of Gamma $k\to\infty$, where $k=\mu/(1+\mu\sigma_{\rm CV}^2)$, since the skewness and excess kurtosis then vanish and the distribution approaches a Gaussian.} but not the qualitative dependence on $n_{\rm field}$ and $\langle N\rangle$. The corresponding standard deviation of the sample variance is therefore
\begin{equation}
\delta s_N^2 \approx \sqrt{{\rm Var}(s_N^2)}
\approx
\sqrt{\frac{2}{n_{\rm field}-1}}\,{\rm Var}(N).
\end{equation}
Propagating this uncertainty into \(\sigma_{\rm CV}^2\) (Equation F2), we get
\begin{equation}
\delta(\sigma_{\rm CV}^2)\approx \frac{\delta s_N^2}{\langle N\rangle^2} = \frac{1}{\langle N\rangle^2}
\sqrt{\frac{2}{n_{\rm field}-1}}\,{\rm Var}(N).
\end{equation}

\noindent Therefore, the approximate signal-to-noise ratio for detecting nonzero cosmic variance is
\begin{equation}
\rm{S/N}(\sigma_{\rm CV}^2) = \frac{\sigma_{\rm CV}^2}{\delta(\sigma_{\rm CV}^2)}
\approx
\sqrt{\frac{n_{\rm field}-1}{2}}\,
\frac{\sigma_{\rm CV}^2}{1/\langle N\rangle+\sigma_{\rm CV}^2},
\end{equation}
where we have used the Equation F1. Finally, because $\delta(\sigma_{\rm CV}^2) = 2\sigma_{\rm CV} \delta(\sigma_{\rm CV})$, we get $\rm S/N(\sigma_{\rm CV}) = 2\,\rm{S/N}(\sigma_{\rm CV}^2)$. 

The expression above shows that, at fixed intrinsic $\sigma_{\rm CV}$, the detectability of cosmic variance increases approximately as $\sqrt{n_{\rm field}}$. This effect is especially important for rare populations, for which $\langle N\rangle$ is small. In that regime, the Poisson term dominates the field-to-field variance, so the uncertainty on the excess-variance estimate can exceed the intrinsic cosmic-variance signal. As a result, many realizations are statistically consistent with zero excess variance, and after imposing the physical condition $\sigma_{\rm CV}\ge 0$, they accumulate at $\sigma_{\rm CV}=0$. The bootstrap distribution therefore develops a pile-up at zero with a tail toward positive values, so summary statistics such as the median can be biased low when the number of sightlines is small. Increasing the number of independent sightlines reduces this effect by shrinking the uncertainty on the measured field-to-field variance. This is consistent with the main result of this work, in which the large number of independent fields is essential for empirically constraining the field-to-field variance of massive quiescent galaxies at $z>3$. This also highlights that if the cosmic variance is high (especially if $\sigma_{\rm CV}>1$), it is possible to detect if from a limited number of sightlines. For example, the fact that \cite{Weibel2025b} manage to detect cosmic variance at $z\sim10$, even though the galaxy number densities are lower than in this work, is fully a result of their galaxy population having an intrinsically higher cosmic variance than the one investigated here.

\end{document}